\def \VersionAuthorforArXiV {}
\ifdefined\VersionAuthorforArXiV
	\newcommand{\arXivVersion}[1]{#1}
	\newcommand{\FinalACMVersion}[1]{}
\else
	\newcommand{\arXivVersion}[1]{}
	\newcommand{\FinalACMVersion}[1]{#1}
\fi

\ifdefined\VersionAuthorforArXiV
	\documentclass[]{article} %
\else
	\PassOptionsToPackage{svgnames,table}{xcolor}
	\documentclass[manuscript, acmsmall, screen=true]{acmart} %
\fi

\usepackage[utf8]{inputenc}
\usepackage[T1]{fontenc}
\usepackage[english]{babel}

\usepackage{csquotes}

\usepackage[ruled,vlined,linesnumbered]{algorithm2e}
	\SetKwInOut{Input}{input}
	\SetKwInOut{Output}{output}

\usepackage{subcaption}

\usepackage{paralist} %

\newenvironment{ienumerate}
	{\begin{inparaenum}[\itshape i\upshape)]}
	{\end{inparaenum}}

\newenvironment{oneenumerate}
	{\ifdefined\VersionLong\begin{enumerate}\else\begin{inparaenum}[1)]\fi}
	{\ifdefined\VersionLong\end{enumerate}\else\end{inparaenum}\fi}

\usepackage{amsmath} %
\arXivVersion{%
	\usepackage{amssymb} %
}
\usepackage{xspace}

\usepackage{graphicx}
\graphicspath{{images}{pictures}}

\usepackage{multirow}

\def \VersionLong {}
\ifdefined \VersionLong
	\newcommand{\LongVersion}[1]{#1}
	\newcommand{\ShortVersion}[1]{}
\else
	\newcommand{\LongVersion}[1]{}
	\newcommand{\ShortVersion}[1]{#1}
\fi

\ifdefined\VersionWithComments
	\usepackage{draftwatermark}
	\SetWatermarkText{draft}
	\SetWatermarkScale{3}
	\SetWatermarkColor[gray]{0.9}
\fi
\ifdefined\LaTeXdiff
	\usepackage{draftwatermark}
	\SetWatermarkText{diff}
	\SetWatermarkScale{3}
	\SetWatermarkColor[gray]{0.9}
\fi
\ifdefined\VersionAuthorforArXiV
	\usepackage[backend=biber,backref=true,style=alphabetic,url=false,doi=true,defernumbers=true,sorting=anyt,maxnames=99]{biblatex} %
	\renewbibmacro*{doi+eprint+url}{%
		\iftoggle{bbx:doi}
			{\color{black!40}\footnotesize\printfield{doi}}
			{}%
		\newunit\newblock
		\iftoggle{bbx:eprint}
			{\usebibmacro{eprint}}
			{}%
		\newunit\newblock
		\iftoggle{bbx:url}
			{\usebibmacro{url+urldate}}
			{}%
	}
	
\fi
\arXivVersion{%
	\usepackage[svgnames,table]{xcolor}
	\definecolor{darkblue}{rgb}{0, 0, 0.7}

	\usepackage[
			pdfauthor={Johan Arcile and Étienne André},%
			pdftitle={Timed automata as a formalism for expressing security: A survey on theory and practice},
			breaklinks  = true,
			colorlinks  = true,
			citecolor   = blue!50!blue,
			linkcolor   = darkblue,
			urlcolor    = blue!50!green,
		]{hyperref}
}

\usepackage[capitalise,english,nameinlink]{cleveref} %
\crefname{line}{\text{line}}{\text{lines}} %

\usepackage{tikz}
\usetikzlibrary{arrows,automata,shapes.gates.logic}
\tikzstyle{every node}=[initial text=]
\tikzstyle{pta}=[auto, ->, >=stealth']
\tikzstyle{location}=[rectangle, rounded corners, minimum size=12pt, draw=black, fill=blue!10, inner sep=2pt]
\tikzstyle{invariant}=[draw=black, dotted, inner sep=1pt, yshift=-20] %
\tikzstyle{final}=[double, fill=blue!50]

\tikzstyle{urgent}=[fill=yellow, thick, dotted] %
\tikzstyle{private}=[fill=red,thick]

\definecolor{coloract}{rgb}{0.50, 0.70, 0.30}
\definecolor{colorclock}{rgb}{0.4, 0.4, 1}
\definecolor{colordisc}{rgb}{1, 0, 1}
\definecolor{colorloc}{rgb}{0.4, 0.4, 0.65}
\definecolor{colorparam}{rgb}{1, 0.6, 0.0}

\definecolor{loccolor1}{rgb}{1, 0.3, 0.3}
\definecolor{loccolor2}{rgb}{0.3, 1, 0.3}
\definecolor{loccolor3}{rgb}{0.3, 0.3, 1}
\definecolor{loccolor4}{rgb}{1, 0.3, 1}
\definecolor{loccolor5}{rgb}{1, 1, 0.3}
\definecolor{loccolor6}{rgb}{0.3, 1, 1}
\definecolor{loccolor7}{rgb}{0.9, 0.6, 0.2}
\definecolor{loccolor8}{rgb}{0.7, 0.4, 1}
\definecolor{loccolor9}{rgb}{0.5, 1, 0.75}
\definecolor{loccolor10}{rgb}{0.8, 0.7, 0.6}
\definecolor{loccolor11}{rgb}{0.6, 0.7, 0.8}
\definecolor{loccolor12}{rgb}{0.2, 0.5, 0.9}
\definecolor{loccolor13}{rgb}{0.5, 0.9, 0.2}
\definecolor{loccolor14}{rgb}{0.9, 0.2, 0.5}
\definecolor{loccolor15}{rgb}{0.7, 0.7, 0.7}
\definecolor{loccolor16}{rgb}{0.8, 0.8, 0.5}

\newcommand{\styleact}[1]{\ensuremath{\textcolor{coloract}{{#1}}}}
\newcommand{\styleclock}[1]{\ensuremath{\textcolor{colorclock}{{#1}}}}
\newcommand{\styledisc}[1]{\ensuremath{\textcolor{colordisc}{\mathrm{#1}}}}

\newcommand{\styleparam}[1]{\ensuremath{\textcolor{colorparam}{#1}}} %

\newcommand{\rowHeader}{\rowcolor{blue!20}}

\newcommand{\cellYes}{\cellcolor{green!40}\textbf{$\surd$}}
\newcommand{\cellNo}{\cellcolor{red!40}\textbf{$\times$}}

\arXivVersion{%
	\usepackage{amsthm}
	\theoremstyle{plain}
	\theoremstyle{definition}
	\newtheorem{definition}{Definition}
	\newtheorem{example}{Example}

	\theoremstyle{remark}

}

\ifdefined\VersionWithComments
	\usepackage[colorinlistoftodos,textsize=footnotesize]{todonotes}
\else
	\usepackage[disable]{todonotes}
\fi

\ifdefined \VersionWithComments
	\newcommand{\todoinline}[1]{\mbox{}{\color{red}{\textbf{TODO}\ifx#1\\\else:\ \fi #1}}} %
\else
	\newcommand{\todoinline}[1]{}
\fi

\usepackage{verbatim} %

\newcommand{\init}{_0}

\newcommand{\A}{\ensuremath{\mathcal{A}}}
\newcommand{\Actions}{\ensuremath{\Sigma}}
\newcommand{\ActionsH}{\ensuremath{\Actions_H}}
\newcommand{\ActionsL}{\ensuremath{\Actions_L}}
\newcommand{\action}{\ensuremath{a}}
\newcommand{\actionSilent}{\ensuremath{\epsilon}}
\newcommand{\actl}{\ensuremath{\styleact{l}}}
\newcommand{\actli}[1]{\ensuremath{\styleact{l_{#1}}}}
\newcommand{\acth}{\ensuremath{\styleact{h}}}
\newcommand{\assign}{\leftarrow}

\newcommand{\Clock}{\mathbb{X}} %
\newcommand{\ClockCard}{H} %
\newcommand{\clock}{x} %
\newcommand{\clockx}{\ensuremath{\styleclock{x}}} %
\newcommand{\clocky}{\ensuremath{\styleclock{y}}} %
\newcommand{\clockval}{\mu} %
\newcommand{\ClocksZero}{\vec{0}}
\newcommand{\compOp}{\bowtie}

\newcommand{\edge}{e}
\newcommand{\Edges}{E}
\newcommand{\EdgesOut}{\ensuremath{\mathit{out}}} %
\newcommand{\longuefleche}[1]{\stackrel{#1}{\longrightarrow}}
\newcommand{\longueflecheRel}[1]{\stackrel{#1}{\mapsto}}

\newcommand{\flecheRel}{{\rightarrow}}

\newcommand{\guard}{g}

\newcommand{\invariant}{I}
\newcommand{\Lg}{\ensuremath{\mathcal{L}}}
\newcommand{\loc}{\ensuremath{\ell}} %
\newcommand{\locinit}{\loc\init}
\newcommand{\Loc}{\mathcal{L}} %

\newcommand{\Param}{P} %
\newcommand{\param}{p} %
\newcommand{\paramp}{\styleparam{p}} %
\newcommand{\probdistr}{\ensuremath{\pi}} %
\newcommand{\pval}{v} %
\newcommand{\resets}{R}

\newcommand{\setQ}{\ensuremath{\mathbb Q}}
\newcommand{\setR}{\ensuremath{\mathbb R}}
\newcommand{\setRgeqzero}{\setR_{\geq 0}}
\newcommand{\setZ}{\ensuremath{\mathbb Z}}
\newcommand{\sinit}{s\init} %
\newcommand{\cstate}{\ensuremath{s}} %
\newcommand{\States}{S} %

\newcommand{\word}{\textcolor{colorok}{w}}

\newcommand{\hide}[2]{\ensuremath{#1_{\setminus #2}}}
\newcommand{\reset}[2]{\ensuremath{[#1]_{#2}}}
\newcommand{\restrict}[2]{\ensuremath{#1_{|#2}}}

\newcommand{\valuate}[2]{\ensuremath{#2(#1)}}

\newcommand{\imitator}{\textsf{IMITATOR}}
\newcommand{\SpaceEx}{\textsf{SpaceEx}}
\newcommand{\uppaal}{\textsc{Uppaal}}
\newcommand{\Verics}{\textsc{VerICS}}
\newcommand{\Yices}{\textsc{Yices2}}

\ifdefined \VersionWithComments
 	\definecolor{colorok}{RGB}{80,80,150}
\else
	\definecolor{colorok}{RGB}{0,0,0}
\fi

\newcommand{\eg}{\textcolor{colorok}{e.\,g.,}\xspace}
\newcommand{\etal}{\textcolor{colorok}{\emph{et al.}}\xspace}
\newcommand{\ie}{\textcolor{colorok}{i.\,e.,}\xspace}
\newcommand{\st}{\textcolor{colorok}{s.t.}\xspace}

\newcommand{\wrt}{\textcolor{colorok}{w.r.t.}\xspace}

\newcommand{\ouracks}{%
	This work is partially supported by the ANR-NRF French-Singaporean research program \href{https://www.loria.science/ProMiS/}{ProMiS} (ANR-19-CE25-0015).
	We would like to thank anonymous reviewers for their useful comments,
	as well as Jaime Arias and Laure Petrucci for a feedback on their recent works.
}

\newcommand{\ourkeywords}{timed automata, cybersecurity, opacity, attack trees, survey}

\newcommand{\ourabstract}{%
	Timed automata are a common formalism for the verification of concurrent systems subject to timing constraints.
	They extend finite-state automata with clocks, that constrain the system behavior in locations, and to take transitions.
	While timed automata were originally designed for \emph{safety} (in the wide sense of correctness \wrt{} a formal property), they were progressively used in a number of works to guarantee \emph{security} properties.
	In this work, we review works studying security properties for timed automata in the last two decades.
	We notably review theoretical works, with a particular focus on opacity, as well as more practical works, with a particular focus on attack trees and their extensions.
	We derive main conclusions concerning open perspectives, as well as tool support.
}

\arXivVersion{
\makeatletter
\def\orcidID#1{\,\smash{\href{https://orcid.org/#1}{\protect\raisebox{%
	+1.25pt%
}{\protect\includegraphics{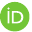}}}}}
\makeatother
}

\FinalACMVersion{%
	\acmJournal{CSUR}
}

\begin{document}
\sloppy

\title{Timed automata as a formalism for expressing security: A survey on theory and practice\arXivVersion{\thanks{%
	This is the author version of the manuscript of the same name published in ACM Computing Surveys.
	The final version is available at \href{https://www.doi.org/10.1145/3534967}{\nolinkurl{doi.org/10.1145/3534967}}.
	\ouracks{}
	}}
}

\ifdefined\VersionAuthorforArXiV
	\author{Johan Arcile\orcidID{0000-0001-9979-3829}
		and
		\'Etienne Andr\'e\orcidID{0000-0001-8473-9555}\\
		Université de Lorraine, CNRS, Inria, LORIA, Nancy, France
	}
	\date{}
\else
	\author{Johan Arcile}
	\orcid{0000-0001-9979-3829}
	\affiliation{%
	\institution{Université de Lorraine, CNRS, Inria, LORIA, F-54000 Nancy, France}
	\city{Nancy}
	\country{France}
	}

	\author{\'Etienne Andr\'e%
	}
	\orcid{0000-0001-8473-9555}
	\affiliation{%
	\institution{Université de Lorraine, CNRS, Inria, LORIA, F-54000 Nancy, France}
	\city{Nancy}
	\country{France}
	}

\begin{abstract}
	\ourabstract{}
\end{abstract}



\fi
\maketitle

\arXivVersion{%
	\newcommand{\keywords}[1]
	{%
		\small\textbf{\textit{Keywords---}} #1
	}

	\begin{abstract}
		\ourabstract{}
	\end{abstract}

}

\todo{This is the version with comments. To disable comments, comment out line~3 in the \LaTeX{} source.}

\ifdefined\VersionWithComments
	\tableofcontents{}
\fi
\section{Introduction}\label{section:introduction}

Numerous critical information systems rely on communication via a shared network, such as the Internet.
Data passing through such networks is often sensitive, and requires secrecy.
If not handled carefully, information such as private data, authentication code, timing information or localization can be accessible to anyone on the network.
This can result in security attacks to retrieve or alter sensitive data \cite{KNNKJ07,HZN09,MPBPPR13}.
To prevent such intrusions, various security methods and protocols have been developed. %
Yet, these security decisions do not always avoid intrusion.

In order to analyze the security of information systems and highlight their weaknesses, technical standards such as FMEA (Failure Mode Effects and Criticality Analysis)~\cite{XTXHZ02,CMFVP06} have been used since the early days of critical information systems.
More structured, model-based approaches have since been explored, such as the ADVISE method~\cite{LFKSM11}, which allows for automatic generation of quantitative metrics, or formalisms such as team automata~\cite{BLP05} and attack trees~\cite{KPS14}, respectively aiming at a formal representation of the behavior of the system and the attacker.
Those approaches allow to check some security properties, such as \emph{authentication} (not being able to lie about oneself), \emph{secrecy} (private information is not visible to public parties), \emph{integrity} (information cannot be altered), and \emph{non-interference} (not being able to deduce another user's actions).

\paragraph{Time and security}
Some security protocols are time-sensitive, by combining concurrency and real-time constraints.
It has been noted in \eg{}~\cite{KPJJ13,BCLR15}
that time is a potential attack vector against otherwise secure systems.
That is, it is possible that a secure system (``non-interferent'') can become insecure (``interferent'') when timing constraints are added~\cite{GMR07}.
In addition, it is possible to correlate the execution time and the value of an encryption key via so-called timing attacks~\cite{Kocher96,FS00,BB07,KPJJ13,BCLR15}.
In 2018, the Spectre vulnerability~\cite{KHFGGHHLMP20} exploited speculative execution to bring secret information into the cache. Subsequently, cache-timing attacks were launched to exfiltrate these secrets.

Timed automata (TAs)~\cite{AD94} are a formalism extending finite-state automata with real-time variables called \emph{clocks}.
TAs were well-studied over the past three decades, and the formalism is supported by a wide range of model checking tools and techniques %
(\eg{} \cite{LPY97,Y97,Fehnker99,NS03,BLR04,MMSYV05,HV06,LSD11,DJLMT15}).
Among them, %
\uppaal{}~\cite{LPY97} is certainly one of the most efficient tools for reachability-based verification, with numerous other extensions.

In the past two decades, a growing number of works considered timed automata in a security context.
Some of these works studied the theory of some security properties (such as non-interference) in the context of TAs.
Some other works used timed automata as a convenient formalism to formally verify some concrete security properties on a given case study, using some model checker (typically \uppaal{}).
Some other works used timed automata as a target formalism from security formalisms, typically attack trees and their extensions, thus allowing to use existing model checkers in order to guarantee these properties.

\paragraph{Related surveys}
To the best of our knowledge, no previous work specifically surveys timed automata in the context of security problems.

Concerning timed automata, Alur and Madhusudan survey in~\cite{AM04} decision problems related to the formalism.
That survey focuses on theoretical aspects of TAs, and is not related to specific applications.
In~\cite{WDR13}, Bin Waez, Dingel and Rudie survey tools and methods using TAs in general; again, no focus is made on security applications.
Also, in~\cite{FC14,KFC17}, Fontana and Cleaveland survey various extensions of TAs, and provide conversion algorithms from these extensions to the original ``vanilla'' TA formalism.
Concerning extensions of timed automata, André surveyed in~\cite{Andre19STTT} decision problems related to various \emph{parametric} extensions of timed automata~\cite{AHV93}.

Regarding applications of formal methods to security issues, Dalal \etal{} benchmark in~\cite{DSHJ10} the performance of state-of-the-art tools for the verification of security protocols, while Avalle \etal{} survey in~\cite{APS14} formal verification techniques for security protocol implementations.
The real-time aspect of systems is not considered.
In~\cite{SN19}, Souri and Norouzi survey formal methods in the context of Internet of Things (IoT) applications, including security aspects.
	Although some of the presented papers use TAs as a modeling formalism, the focus of the survey is on the applications rather than the methods.
Finally, in~\cite{WAFP19} (which extends~\cite{KPS14}), Widel \etal{} survey methods using attack trees, a formalism used in the context of security modeling.
Among those methods, TAs are present as a way to model such trees, but the focus is not specifically made on TAs.
\paragraph{Contribution}
We provide here a survey of the use of timed automata and their extensions in the context of security analyses.
We believe surveying works relating timed automata and security can help readers to understand the state-of-the-art of both the theory (``what security problems are decidable for timed automata and their subclasses'') and the practice (``what can be practically achieved, using what software'').
We emphasize on open problems, and on practical TA model checkers that can be used for solving security problems.

Somehow surprisingly, while timed automata seem to be an interesting formalism to model interactions between time and concurrency in the framework of security property, the number of such works is ``reasonable'' (very roughly a hundred or so).
Therefore, we aim at a sense of near-completeness in this survey, so as to provide readers with a complete overview of the works using timed automata as an approach towards verifying security properties.
In addition, we use a limited set of TA examples on which we illustrate various (theoretical) notions from the literature, so as to exemplify the (sometimes subtle) differences between these notions in a unified manner.

When surveying these works, two large research directions emerged:
\begin{enumerate}
	\item the works studying theoretical aspects, notably linked to the decidability of various opacity and non-interference properties, on various subclasses of TAs;
	\item the works translating increasingly larger extensions of attack trees into (extensions of) TAs.
\end{enumerate}
These works are therefore surveyed in two dedicated sections.
Most remaining works concern applications to various areas: verification of protocols, of controllers, formalization of RBAC (role-based access control) models, etc.

\paragraph{Methodology}
We collected works using different methods:
\begin{ienumerate}%
	\item performing search engine queries, typically \href{https://dblp.uni-trier.de/}{DBLP} and \href{https://scholar.google.com/}{Google Scholar},
	\item collecting relevant citations to our own works, and
	\item collecting relevant citations from and to all the aforementioned works. %
\end{ienumerate}%
We aim at a (near-)completeness, %
with the goal to be as complete as possible regarding works in the domain of timed automata and security.

\paragraph{Outline of the manuscript}
\cref{section:timed_automata} gives a formal definition of timed automata syntax and semantics.
Some extensions of TAs and related tools mentioned in the survey are also presented.
We survey specifically works studying timed automata and non-interference in \cref{section:non-interference}.
We then survey more practical methods concerning translations of attack trees and their extensions into timed automata, in \cref{section:attacktrees}.
We then review in \cref{section:applications} remaining works, that all address various application domains in the area of cyber-physical systems, notably approaches related to formalizing RBAC models, verifying controllers and detecting attacks on communication protocols.
We give some general perspectives in \cref{section:conclusions}.

\section{Timed Automata}\label{section:timed_automata}

Timed automata (TAs)~\cite{AD94} are a common formalism to model concurrent timed systems, and formally verify them against temporal (``order of the events'') and timed (``relative or absolute time at which events occur'') properties.
Among other advantages, TAs can express both a \emph{continuous} representation of time, and \emph{concurrency}.
Notably, checking TAs against timed properties expressed in TCTL (timed computation tree logic)~\cite{ACD93}, a well-known extension of CTL~\cite{CE82}, is decidable~\cite{AD94}---which includes reachability properties.

Syntactically, a TA is a directed and connected graph, featuring an initial vertex, and extended with \emph{clocks}, \ie{} real-valued variables that evolve at the same rate.
Vertices are called \emph{locations} and are associated with \emph{invariants}.
Invariants are constraints on clocks, authorizing entry into a location or forcing its exit.
The edges are annotated with \emph{guards}, along which clocks can be reset to~0.

We recall the formal definition of both the syntax and semantics of TAs; we also review some extensions of TAs mentioned in this manuscript.

\subsection{Clock guards}

We assume a set~$\Clock = \{ \clock_1, \dots, \clock_\ClockCard \} $ of \emph{clocks}, \ie{} real-valued variables that evolve at the same rate.
A clock valuation is a function
$\clockval : \Clock \rightarrow \setRgeqzero$.
We write $\ClocksZero$ for the clock valuation assigning $0$ to all clocks.
Given $d \in \setRgeqzero$, $\clockval + d$ denotes the valuation \st{} $(\clockval + d)(\clock) = \clockval(\clock) + d$, for all $\clock \in \Clock$.
Given $\resets \subseteq \Clock$, we define the \emph{reset} of a valuation~$\clockval$, denoted by $\reset{\clockval}{\resets}$, as follows: $\reset{\clockval}{\resets}(\clock) = 0$ if $\clock \in \resets$, and $\reset{\clockval}{\resets}(\clock)=\clockval(\clock)$ otherwise.

We assume ${\compOp} \in \{<, \leq, =, \geq, >\}$.
A clock guard~$\guard$ is a constraint over $\Clock$ defined by a conjunction of inequalities of the form
$\clock \compOp c$, with
	$c \in \setZ$.\footnote{%
	Several syntax variations exist in the literature, notably \emph{diagonal constraints} of the form $\clock - \clock' \compOp c$.
} %
Given~$\guard$, we write~$\clockval \models \guard$ if %
the expression obtained by replacing each~$\clock$ with~$\clockval(\clock)$ in~$\guard$ evaluates to true.

\subsection{Timed automata}
\begin{definition}[timed automaton]\label{def:TA}
	A \emph{timed automaton} (TA) $\A$ is a tuple \mbox{$\A = (\Actions, \Loc, \locinit, \Clock, \invariant, \Edges)$}, where:
	\begin{enumerate}
		\item $\Actions$ is a finite set of actions,
		\item $\Loc$ is a finite set of locations,
		\item $\locinit \in \Loc$ is the initial location,
		\item $\Clock$ is a finite set of clocks,
		\item $\invariant$ is the invariant, assigning to every $\loc\in \Loc$ a clock guard $\invariant(\loc)$,
		\item $\Edges$ is a finite set of edges  $\edge = (\loc,\guard,\action,\resets,\loc')$
		where~$\loc,\loc'\in \Loc$ are the source and target locations, $\action \in \Actions$, $\resets\subseteq \Clock$ is a set of clocks to be reset, and $\guard$ is a clock guard.
	\end{enumerate}
\end{definition}
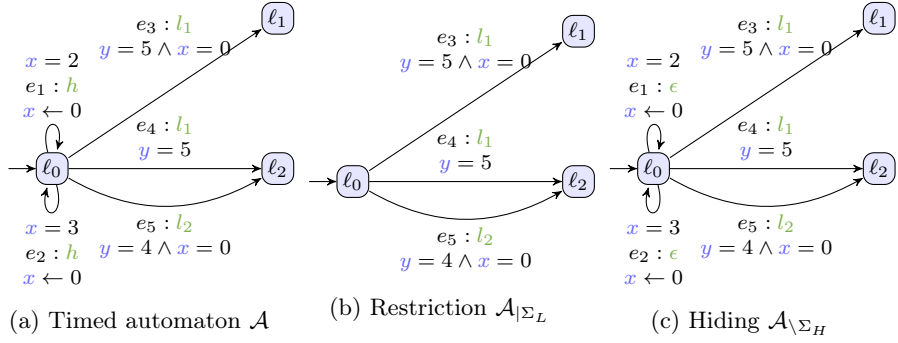
\begin{figure}[tb]

\begin{subfigure}{.32\textwidth}
	\centering
	 \footnotesize
	\begin{tikzpicture}[pta, scale=1, xscale=1, yscale=1]

		\node[location, initial] at (0, 0) (l0) {$\loc_0$};

		\node[location] at (3, 2) (l1) {$\loc_1$};

		\node[location] at (3, 0) (l2) {$\loc_2$};

		\path (l0) edge[loop above] node[above,align=center]{$\clockx = 2$ \\ $\edge_1 : \acth$ \\ $ \clockx \assign 0 $} (l0);
		\path (l0) edge[loop below] node[below,align=center]{$\clockx = 3$ \\ $\edge_2 : \acth$ \\ $ \clockx \assign 0 $} (l0);

		\path (l0) edge node[above,yshift=10,align=center]{$\edge_3 : \actli{1}$ \\ $ \clocky = 5 \land \clockx = 0 $} (l1);
		\path (l0) edge node[above,align=center]{$\edge_4 : \actli{1}$ \\ $ \clocky = 5 $} (l2);
		\path (l0) edge[bend right] node[below,align=center]{$\edge_5 : \actli{2}$ \\ $ \clocky = 4 \land \clockx = 0 $} (l2);

	\end{tikzpicture}
	\caption{Timed automaton $\A$}
	\label{figure:example-n-non-interference}
\end{subfigure}
\begin{subfigure}{.32\textwidth}
	\centering
	 \footnotesize
	\begin{tikzpicture}[pta, scale=1, xscale=1, yscale=1]

		\node[location, initial] at (0, 0) (l0) {$\loc_0$};

		\node[location] at (3, 2) (l1) {$\loc_1$};

		\node[location] at (3, 0) (l2) {$\loc_2$};

		\path (l0) edge node[above,yshift=10,align=center]{$\edge_3 : \actli{1}$ \\ $ \clocky = 5 \land \clockx = 0 $} (l1);
		\path (l0) edge node[above,align=center]{$\edge_4 : \actli{1}$ \\ $ \clocky = 5 $} (l2);
		\path (l0) edge[bend right] node[below,align=center]{$\edge_5 : \actli{2}$ \\ $ \clocky = 4 \land \clockx = 0 $} (l2);

	\end{tikzpicture}
	\caption{Restriction $\restrict{\A}{\ActionsL}$}
	\label{figure:example-restriction}
\end{subfigure}
\begin{subfigure}{.32\textwidth}
	\centering
	 \footnotesize
	\begin{tikzpicture}[pta, scale=1, xscale=1, yscale=1]

		\node[location, initial] at (0, 0) (l0) {$\loc_0$};

		\node[location] at (3, 2) (l1) {$\loc_1$};

		\node[location] at (3, 0) (l2) {$\loc_2$};

		\path (l0) edge[loop above] node[above,align=center]{$\clockx = 2$ \\ $\edge_1 : \styleact{\actionSilent}$ \\ $ \clockx \assign 0 $} (l0);
		\path (l0) edge[loop below] node[below,align=center]{$\clockx = 3$ \\ $\edge_2 : \styleact{\actionSilent}$ \\ $ \clockx \assign 0 $} (l0);

		\path (l0) edge node[above,yshift=10,align=center]{$\edge_3 : \actli{1}$ \\ $ \clocky = 5 \land \clockx = 0 $} (l1);
		\path (l0) edge node[above,align=center]{$\edge_4 : \actli{1}$ \\ $ \clocky = 5 $} (l2);
		\path (l0) edge[bend right] node[below,align=center]{$\edge_5 : \actli{2}$ \\ $ \clocky = 4 \land \clockx = 0 $} (l2);

	\end{tikzpicture}
	\caption{Hiding $\hide{\A}{\ActionsH}$}
	\label{figure:example-hiding}
\end{subfigure}

\caption{Illustrating restriction and hiding}
\label{figure:restriction-and-hiding}

\end{figure}
\begin{example}\label{example:TA}
	Consider the TA in \cref{figure:example-n-non-interference}, containing two clocks~$\clockx$ and~$\clocky$.
	This TA features no invariant.
	For each transition, we denote vertically (from top to bottom) first the guard (if any), then the action, and finally the clocks resets (if any).
	In \cref{figure:restriction-and-hiding}, we also give the edge \emph{name} together with the action (of the form ``$\edge : \action$'' where $\edge$ is the edge name, and $\action$ is the action).
	The edge ``names'' do not strictly speaking belong to the TA, but they will ease describing our subsequent examples using \cref{figure:restriction-and-hiding}.
	
	For example, in \cref{figure:example-n-non-interference}, the self-loop from~$\loc_0$ to~$\loc_0$ via action~$\acth$ is guarded by clock guard ``$\clockx = 2$'', and resets clock~$\clockx$.
	The name of this edge is~$\edge_1$.
	
	From the initial location $\loc_0$ in \cref{figure:example-n-non-interference}, it is possible to reach either~$\loc_1$ or~$\loc_2$.
	Reaching~$\loc_2$ via~$\actli{1}$ can be done by simply waiting 5 time units (guard ``$\clocky = 5$''), while reaching~$\loc_1$ via~$\actli{1}$ or~$\loc_2$ via~$\actli{2}$ is more involved, as (assuming both clocks are initially~0) the TA will need to take several times the self-loop on~$\loc_0$ resetting~$\clockx$ when some guard ($\clockx = 2$ or $\clockx = 3$) is satisfied, while keeping~$\clocky$ unchanged, until eventually the guard ``$\clocky = 5 \land \clockx = 0$'' is satisfied.
\end{example}
\subsubsection{Syntactical transformations}

Sometimes, and notably in works related to \emph{non-interference} or \emph{opacity}, we will assume that actions are partitioned into a set of \emph{low-level actions} $\ActionsL$ and a set of \emph{high-level actions} $\ActionsH$.
The \emph{restriction} of a TA to low-level actions discards all edges labeled with high-level actions.

\begin{definition}[restriction]
	Let $\A = (\Actions, \Loc, \locinit, \Clock, \invariant, \Edges)$ be a TA with $\Actions = \ActionsL \uplus \ActionsH$ ($\uplus$ denotes disjoint union).
	The \emph{restriction of~$\A$ to low-level actions}, denoted by~$\restrict{\A}{\ActionsL}$,
	is defined as the TA identical to $\A$ except that any edge of the form $(\loc,\guard,\action,\resets,\loc')$ with $\action \notin \ActionsL$ (\ie{} $\action \in \ActionsH$) is discarded.
\end{definition}

The hiding of high-level actions in a TA replaces all edges labeled with high-level actions with unobservable transitions.

\begin{definition}[hiding]\label{definition:hiding}
	Let $\A = (\Actions, \Loc, \locinit, \Clock, \invariant, \Edges)$ be a TA with $\Actions = \ActionsL \uplus \ActionsH$.
	The \emph{hiding of high-level actions in~$\A$}, denoted by~$\hide{\A}{\ActionsH}$,
	is defined as the TA identical to $\A$ except that any edge of the form $(\loc,\guard,\action,\resets,\loc')$ with $\action \in \ActionsH$ is replaced with an edge $(\loc,\guard,\actionSilent,\resets,\loc')$,
	where $\actionSilent$ is the special silent (unobservable) action.
\end{definition}
\begin{example}
	Consider again the TA~$\A$ in \cref{figure:example-n-non-interference}.
	Assume $\ActionsL = \{ \actli{1}, \actli{2} \}$ and $\ActionsH = \{ \acth \}$.
	Then the restriction of~$\A$ to~$\ActionsL$ is given in \cref{figure:example-restriction}.
	The hiding of~$\ActionsH$ in~$\A$ is given in \cref{figure:example-hiding}.
\end{example}
\subsubsection{Concrete semantics of TAs}
We recall below the concrete semantics of TAs.

\begin{definition}[Semantics of a TA]
	Given a TA $\A = (\Actions, \Loc, \locinit, \Clock,  \invariant, \Edges)$,
	the semantics of $\A$ is given by the timed transition system (TTS) $(\States, \sinit, \flecheRel)$, with
	\begin{itemize}
		\item $\States = \{ (\loc, \clockval) \in \Loc \times \setRgeqzero^\ClockCard \mid \clockval \models \invariant(\loc) \}$,
		\item $\sinit = (\locinit, \ClocksZero) $,
		\item  $\flecheRel$ consists of the discrete and (continuous) delay transition relations:
		\begin{enumerate}
			\item discrete transitions: $(\loc,\clockval) \longueflecheRel{\edge} (\loc',\clockval')$, %
				if $(\loc, \clockval) , (\loc',\clockval') \in \States$, and there exists $\edge = (\loc,\guard,\action,\resets,\loc') \in \Edges$, such that $\clockval'= \reset{\clockval}{\resets}$, and $\clockval \models \guard$.

			\item delay transitions: $(\loc,\clockval) \longueflecheRel{d} (\loc, \clockval+d)$, with $d \in \setRgeqzero$, if $\forall d' \in [0, d], (\loc, \clockval+d') \in \States$.
		\end{enumerate}
	\end{itemize}
\end{definition}

Moreover we write $(\loc, \clockval)\longuefleche{(d, \edge)} (\loc',\clockval')$ for a combination of a delay and discrete transition if
		$\exists  \clockval'' :  (\loc,\clockval) \longueflecheRel{d} (\loc,\clockval'') \longueflecheRel{\edge} (\loc',\clockval')$.

Given a TA~$\A$ with concrete semantics $(\States, \sinit, \flecheRel)$, we refer to the states of~$\States$ as the \emph{concrete states} of~$\A$.
A \emph{concrete run} of~$\A$ is an alternating sequence of concrete states of $\A$ and pairs of delays and edges starting from the initial state $\sinit$ of the form
$\cstate_0, (d_0, \edge_0), \cstate_1, \cdots$
with
$i = 0, 1, \dots$, $\edge_i \in \Edges$, $d_i \in \setRgeqzero$ and
	$\cstate_i \longuefleche{(d_i, \edge_i)} \cstate_{i+1}$.

Given a state~$\cstate=(\loc, \clockval)$, we say that $\cstate$ is \emph{reachable} in~$\A$ if $\cstate$ appears in a run of~$\A$;
and by extension, we say that $\loc$ is reachable in~$\A$. %

Given a concrete run $(\loc_0, \clockval_0), (d_0, \edge_0), (\loc_1, \clockval_1), \cdots$, the associated \emph{timed word} is
	$(\action_0, \tau_0), (\action_1, \tau_1), \cdots$, where $\action_i$ is the action of edge~$\edge_{i}$, and $\tau_i = \sum_{0 \leq j \leq i} d_j$, for $i = 1, 2 \cdots$.
The timed language of~$\A$, denoted by $\Lg(\A)$ is the set of timed words associated with all concrete runs of~$\A$.\footnote{%
	Several works related to TAs add a notion of \emph{acceptance}, \ie{} the run must end in a \emph{final} location to be accepting and be part of the language.
}
Silent actions ($\actionSilent$) are omitted from the timed words and the subsequent timed language.

\begin{example}
	Let us first come back to the TA in \cref{figure:example-n-non-interference}.
	A possible run for this TA is
		$(\loc_0, (0, 0))
			\longuefleche{(2, \edge_1)}
		(\loc_0, (0,2))
			\longuefleche{(3, \edge_2)}
		(\loc_0, (0,5))
			\longuefleche{(0, \edge_3)}
		(\loc_1, (0,5))
	$.
	Note that, as an abuse of notation, we write $(\loc_0, (0, 2))$ for $(\loc_0, \clockval_0)$ such that $\clockval_0(\clockx) = 0$ and $\clockval_0(\clocky) = 2$.
	
	Since $\loc_1$ belongs to a concrete run of~$\A$, $\loc_1$ is reachable in~$\A$.
	In addition, the timed word associated to the aforementioned run is
	$(\acth, 2), (\acth, 5), (\actli{1}, 5)$.

	Let us now consider the TA in \cref{figure:example-restriction}.
	The timed word $(\acth, 2), (\acth, 5), (\actli{1}, 5)$ is \emph{not} part of this TA, and in fact, it can be shown that $\loc_1$ is unreachable, \ie{} it cannot be reached along any run.
	As the only edge that can be taken is~$\edge_4$ (to~$\loc_2$), the only concrete run is
		$(\loc_0, (0, 0))
			\longuefleche{(5, \edge_4)}
		(\loc_0, (5, 5))$
	and the timed language is exactly $\{ (\actli{1}, 5) \}$.
	
	Let us now consider the TA in \cref{figure:example-hiding}.
	The timed language of this TA is made of two timed words of length~1: 
	\( \{ (\actli{1}, 5) , (\actli{2}, 4) \} \)
	Observe that silent actions are omitted.
	Also, the first timed word $(\actli{1}, 5)$ can correspond to either a run reaching~$\loc_1$ (via~$\edge_3$), or a run reaching~$\loc_2$ (via~$\edge_4$). 
\end{example}
\subsection{Extensions of timed automata}

In the following, we briefly review two extensions of timed automata, both being considered relatively often when discussing security issues.

\subsubsection{Probabilistic timed automata}

Probabilistic timed automata extend TAs with probabilities along edges~\cite{GJ95,KNSS02}.

A probabilistic TA %
is a tuple \mbox{$\A = (\Actions, \Loc, \locinit, \Clock, \invariant, \Edges, \probdistr)$}, where $\probdistr : \Edges \rightarrow [0,1]$ is a probability function.
For $\edge \in \Edges$, $\probdistr(\edge)$ is the probability of taking edge~$\edge$.
Let $\EdgesOut(\loc) \subseteq \Edges$ be the set of outgoing edges from location~$\loc$;
then:
\[\forall \loc \in \Loc : \sum_{\edge \in \EdgesOut(\loc)} \probdistr(\edge) = 1\text{.}\]

\subsubsection{Parametric timed automata}

Parametric TA extend TAs with timing parameters that can be used instead of integer constants in guards and invariants~\cite{AHV93}.
These parameters are in general rational-valued, even though some restrictions may exist (see a survey in~\cite{Andre19STTT}).

Formally, given a set of timing parameters~$\Param$,
a parametric clock guard~$\guard$ is a constraint over $\Clock \cup \Param$ defined by a conjunction of inequalities of the form
$\clock \compOp \param$
or
$\clock \compOp c$,
with
	$\clock \in \Clock$,
	$\param \in \Param$,
	and
	$c \in \setZ$.

A parametric TA is a tuple $\A = (\Actions, \Loc, \locinit, \Clock, \Param, \invariant, \Edges)$,
	where $\invariant(\loc)$ assigns to location~$\loc$ a parametric clock guard,
	and $\Edges$ is a set of edges $\edge = (\loc,\guard,\action,\resets,\loc')$, where $\guard$ is a parametric clock guard.

Given a parameter valuation $\pval : \Param \to \setQ$, we denote by $\valuate{\A}{\pval}$ the non-parametric structure where each occurrence of a parameter $\param_i$ is replaced with~$\pval(\param_i)$.
This structure is usually\footnote{%
	With some assumptions such as $\pval(\param) \in \setZ$ for all~$\param$, or assuming an appropriate rescaling of constants.
} a \emph{timed automaton} as defined in~\cite{AD94}.

\begin{example}
	Consider the PTA~$\A$ in \cref{figure:example-pn-non-interference} (page~\pageref{figure:example-pn-non-interference}), featuring a single parameter~$\param$.
 	Given $\pval$ such that $\pval(\param) = 5$, $\valuate{\A}{\pval}$ corresponds to the TA in \cref{figure:example-n-non-interference}.
\end{example}
\subsection{Model checking tools for (extensions of) timed automata}
Model checking~\cite{CGP01} involves building a (usually finite) model of a system, then checking whether a given property holds on this model by traversing (a portion of) the (finite or infinite) set of the reachable states.
Model checking has the advantages of being fully automatic, and of usually generating concrete witnesses (``traces'') in the case of a successful reachability property.
However, it suffers from the infamous state space explosion problem.
To cope with this problem, model checking tools often rely on specific abstractions that reduce the size of the state space, \ie{} the number of states, (usually) without losing critical information.

The most frequently used model checker among the works surveyed here is \uppaal~\cite{LPY97}, a verification tool supporting an extension of TAs that has been used by the community since the early 2000's.
\uppaal{} extends TAs with features such as integer variables, data structures, functions and synchronization channels.
The tool allows to verify a subset of TCTL (mainly reducing to reachability), and generates concrete execution traces in case of successful reachability.

One of the many extensions of \uppaal{}, \uppaal{}-SMC~\cite{DLLMP15}, is mentioned a few times in this survey.
\uppaal{}-SMC supports probabilistic TAs, and uses statistical model checking (SMC) to avoid the exhaustive exploration of the state space.
Basically, SMC consists in a statistical approach that allows to know whether a given property is true with a given degree of confidence.
Although it lacks exhaustiveness, it is generally well-suited for checking stochastic properties.

\imitator{}~\cite{Andre21} is mentioned in this survey when reviewing works in which some timing parameters are unknown.
\imitator{} takes as input extensions of TAs with \emph{timing parameters}, and performs parametric timed model checking: rather than answering a binary answer (``yes''/``no'') to a given property, it synthesizes a set of parameter valuations enforcing the property.

\Verics{}~\cite{KNNPPSWZ08} is used in the context of communication protocols.
It is a SAT-based verification tool (\ie{} the model and the property are translated into a Boolean formula that holds if the property holds) that supports TAs.
It allows bounded model checking (BMC~\cite{BCCSZ03}).

Finally, the following other model checking tools are mentioned in some of the works surveyed in the following: %
	Kronos~\cite{Y97}, MiniSAT~\cite{SE05}, Modest~\cite{BDHK06}, PAT~\cite{LSD11}, \uppaal{}-CORA\footnote{%
		\url{https://docs.uppaal.org/extensions/cora/}
	}, and \uppaal{}-Stratego~\cite{DJLMT15}.

\section{Non-interference}\label{section:non-interference}

In this section, we survey methods related to the timed modeling of information flow, allowing to verify non-interference properties.
Non-interference concerns systems where different levels of security are present in the system (typically, normal users and administrators).
Such systems satisfy non-interference if there is no information flow from the high security level to the low security level.
Intuitively, actions taken on the higher security level (including the absence of actions) does not impact in any way the lower level.
Equivalently, there is no way for a low level user to infer information on the higher level.

Non-interference can become more challenging when timing information are involved.
Without the non-interference property, systems are vulnerable to attacks that exploit the timing of message flows to infer information.
Typically, it is possible to infer the content of some memory space from the access times of a cryptographic module.
Such attacks are surveyed in~\cite{BGN17}.

Time particularly influences the non-interference of a system.
It has been shown in~\cite{GMR07} that a non-interferent system can become interferent when timing constraints are added.

In this entire section, the attacker (if any) has access to the entire system model, and can observe at runtime some partial behavior (for example, low-level actions together with their timing).

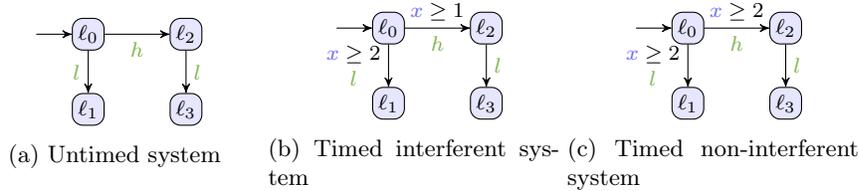
\begin{figure}[tb]

	\centering
	 \footnotesize

\begin{subfigure}{.32\textwidth}
\centering
	\begin{tikzpicture}[pta, scale=1, xscale=1.3, yscale=1]

		\node[location, initial] at (0, 0) (l0) {$\loc_0$};
		\node[location] at (0, -1) (l1) {$\loc_1$};
		\node[location] at (1, 0) (l2) {$\loc_2$};
		\node[location] at (1, -1) (l3) {$\loc_3$};
		
		\path (l0) edge node[align=center,left]{$\actl{}$} (l1);
		\path (l0) edge node[above,align=center]{}node[below]{$\acth{}$} (l2);
		\path (l2) edge node[align=center]{$\actl{}$} (l3);

	\end{tikzpicture}
	\caption{Untimed system}
	\label{figure:noninterference:untimed}
\end{subfigure}
\begin{subfigure}{.32\textwidth}
  \centering
	\begin{tikzpicture}[pta, scale=1, xscale=1.3, yscale=1]

		\node[location, initial] at (0, 0) (l0) {$\loc_0$};
		\node[location] at (0, -1) (l1) {$\loc_1$};
		\node[location] at (1, 0) (l2) {$\loc_2$};
		\node[location] at (1, -1) (l3) {$\loc_3$};
		
		\path (l0) edge node[align=center,left]{$\styleclock{x} \geq 2$\\$\actl{}$} (l1);
		\path (l0) edge node[above,align=center]{$\styleclock{x} \geq 1$}node[below]{$\acth{}$} (l2);
		\path (l2) edge node[align=center]{$\actl{}$} (l3);

	\end{tikzpicture}
	\caption{Timed interferent system}
	\label{figure:noninterference:no}
\end{subfigure}
\begin{subfigure}{.32\textwidth}
  \centering
	\begin{tikzpicture}[pta, scale=1, xscale=1.3, yscale=1]

		\node[location, initial] at (0, 0) (l0) {$\loc_0$};
		\node[location] at (0, -1) (l1) {$\loc_1$};
		\node[location] at (1, 0) (l2) {$\loc_2$};
		\node[location] at (1, -1) (l3) {$\loc_3$};
		
		\path (l0) edge node[align=center,left]{$\styleclock{x} \geq 2$\\$\actl{}$} (l1);
		\path (l0) edge node[above,align=center]{$\styleclock{x} \geq 2$}node[below]{$\acth{}$} (l2);
		\path (l2) edge node[align=center]{$\actl{}$} (l3);

	\end{tikzpicture}
	
	\caption{Timed non-interferent system}
	\label{figure:noninterference:yes}
\end{subfigure}%
	\caption{Illustrating non-interference and timed automata}
	\label{figure:noninterference}
\end{figure}
\begin{example}
Let us informally illustrate the notion of non-interference for timed systems.
Consider the TAs in \cref{figure:noninterference}, featuring a single clock~$\clock$ and two actions: $\actl{}$ is a low-level action, while $\acth{}$ is a high-level action.
When referring to non-interference as a notion related to the observable timed language (\ie{} only the low-level actions, with their timestamp, are observable to an external observer), then the TA in \cref{figure:noninterference:no} is \emph{not} non-interferent.
Indeed, if an attacker (that has access to the TA model) observes $\actl{}$ at time~$1.5$, then they can deduce that the edge from~$\loc_0$ to~$\loc_2$ and then to~$\loc_3$ was necessarily taken, and therefore $\acth{}$ has necessarily occurred.

In contrast, the TA in \cref{figure:noninterference:yes} is non-interferent: it is not possible to deduce information on the occurrence of high-level actions by only observing low-level actions (with their timestamp).

As a side-note on this example, also observe that, as~\cite{GMR07} mentioned, \cref{figure:noninterference:untimed} is an untimed non-interferent automaton, but adding timing constraints (\cref{figure:noninterference:no}) makes it interferent.
\end{example}
\paragraph{Outline of the section}
In this section, we review various notions of non-interference for timed automata, namely
$n$-non-interference (\cref{ss:nnoninterference}),
non-interference and simulation (\cref{ss:noninterference+sim}),
non-interference and probabilities (\cref{ss:noninterference+prob}),
timed opacity (\cref{ss:timed-opacity}),
and
non-interference and information flow (\cref{ss:noninterference+flow}).
We also provide a local summary and discussion in \cref{ss:noninterference:discussion}.

\subsection{$n$-non-interference}\label{ss:nnoninterference}

A first occurrence of the use of TAs in the context of non-interference is found in~\cite{FGLMMTT01}, where Focardi \etal{} study a known privacy breach on websites, where a timed attack allows a malicious website to determine whether another unrelated website is present in the user's Web-browsing history.
The website, the cache, and the user requests are modeled using TAs, where actions are divided between high-level (the user) and low-level (the malicious website) ones.
A model is obtained by doing the parallel composition of these automata; then, the authors deduce whether there is a possibility of information flow from high-level actions to low-level ones.
The results highlight that the attack is possible only if the requests from the user are separated by less than a certain amount of time.
Those results are compared with both symbolic model checking based on binary decision diagram~\cite{BCCSZ03}, and with process algebras~\cite{FGM00}.
The TAs approach has the advantage of using continuous time, but verifying properties takes longer.

\paragraph{Timed $n$-non-interference}
Barbuti \etal{} are the first to propose in~\cite{BDST02} a method using TAs to model and analyze timed non-interference.
As usual, the set of actions in the considered TAs is supposed to be partitioned into high-level and low-level actions.
Non-interference is defined in regard to a minimum delay~$n$ between consecutive high-level actions.
A given timed automaton $\A$ is $n$-non-interfering if the language of~$\A$ %
	subject to the hiding of its high-level actions~$\ActionsH$ (\ie{} $\hide{\A}{\ActionsH}$, see \cref{definition:hiding})
is equal to the language of the TA based on~$\A$ but where high-level actions are allowed if being separated by at least $n$ time units.
This latter condition is ensured by taking the parallel composition of~$\A$ with a gadget TA ensuring that any two high-level actions are separated by at least $n$ time units.
Formal details are given about the transformation of the TA, along with an example on a simplified airplane controller.
However, checking equivalence between two timed languages is undecidable for general TAs~\cite{AD94}, and therefore the practical side of checking $n$-non-interference is not addressed in this work.

In~\cite{BT03}, the authors adapt the approach from~\cite{BDST02} with a different notion of non-interference.
The difference is that the equivalence is now checked on the set of reachable locations (called ``states'' in~\cite{BT03}) rather than on the timed language.
This equivalence is decidable as it reduces to the reachability problem for TAs, which is decidable~\cite{AD94}.
An example of this new method is illustrated on Fischer's mutual exclusion protocol~\cite{L87}.
Two processes and an intruder able to rewrite a shared variable are modeled using TAs.
With this new definition, proving $n$-non-interference reduces to checking whether locations that are not reachable without the intruder are also not reachable with the intruder but with actions of the two processes separated by time intervals of at least $n$~time units.
It is then possible to infer a minimum waiting time between two high-level actions of the process, such that the system verifies non-interference.

\begin{example}\label{example:timednnoninterference}
	Consider again the TA in \cref{figure:example-n-non-interference}, featuring two low-level actions~$\actli{1}$ and~$\actli{2}$, and one high-level action~$\acth$.
	In absence of high-level actions (\ie{} restricted to low-level actions, see \cref{figure:example-restriction}), note that, since $\clockx$ is never reset, only the edge~$\edge_4$ (from~$\loc_0$ to~$\loc_2$ with action~$\actli{1}$) can be taken.
	Therefore, the timed language of this TA restricted to low-level actions is $\{ (\actli{1}, 5) \}$ and the only reachable locations are $\{\loc_0, \loc_2\}$.
	However, when one allows high-level actions (\cref{figure:example-n-non-interference}), the transitions resetting~$\clockx$ (edges $\edge_1$ and~$\edge_2$) can be taken every 2 or 3~time units, therefore potentially modifying the observable behavior.

	Let us first illustrate non-interference on timed language (\cite{BDST02}).
	Recall that the only timed word in the absence of~$\acth$ (\ie{} in the TA restricted to low-level actions, in \cref{figure:example-restriction}) is $(\actli{1}, 5)$.
	Note that taking the edge~$\edge_3$ to~$\loc_1$ does not change this timed word.
	The only way to obtain a different timed word is to take the lower transition~$\edge_5$ from~$\loc_0$ to~$\loc_2$, labelled with~$\actli{2}$.
	This transition can only be taken after taking twice the lower self-loop ($\edge_2$) below~$\loc_0$---which requires a minimum delay of $2$ time units.
	Therefore, the system is not 2-non-interferent but is $n$-non-interferent for any $n \geq 3$.
	Note that~\cite{BDST02} does not perform a \emph{parameterized} verification (for any~$n$), but requires a different check for any valuation of~$n$.

	Concerning non-interference \wrt{} the reachability of locations (\cite{BT03}), the only way to obtain a different set of reachable locations is to take the transition~$\edge_3$ to~$\loc_1$.
	To do so, one needs to take the self-loop~$\edge_1$ above~$\loc_0$ (\ie{} after 2 time units) and then the self-loop~$\edge_2$ below~$\loc_0$ (\ie{} after 3 time units)---or vice-versa---so that eventually the guard $\clocky = 5 \land \clockx = 0$ is satisfied.
	This is possible whenever the frequency of the high-level action (\ie{} $\acth$) is high enough: for $n = 1$ or $n = 2$, this is possible.
	And therefore, the system is neither 1-non-interferent nor 2-non-interferent.
	However, the system is 3-non-interferent because, even though the high-level actions are allowed, there is no way to perform twice $\acth$ within 5 time units (required by the guard $\clocky = 5$), with a minimum duration of 3 time units in between (required by the 3-non-interference property).
	Again, a \emph{parameterized} verification (for any~$n$) is not discussed in~\cite{BT03} (it will be addressed in~\cite{AK20}, see below).
\end{example}
\paragraph{Generalized abstract non-interference}
In~\cite{GM05}, Giacobazzi and Mastroeni propose a notion of generalized abstract non-interference (GANI), extending abstract non-interference from~\cite{GM04}.
This notion applies to any kind of finite-state machine models.
An abstraction of the observable information flow and of a restricted information flow (hiding what must be kept private) must be given.
The system verifies GANI if the observation of the attacker on the observable information flow is equivalent to its observation on the restricted information flow.
The authors prove that, in the context of TAs, GANI includes the $n$-non-interference from~\cite{BDST02}.
\paragraph{Parametric $n$-non-interference}
André and Kryukov~\cite{AK20} later extend the decidable $n$-non-interference on reachable locations from~\cite{BT03} to the formalism of parametric timed automata~\cite{AHV93}.
They define the $n$-location-non-interference synthesis problem on parametric TAs as the computation of all timed intervals ($n$) and parameter values for which the system is non-interferent with respect to the set of reachable locations.
The synthesis is done using the parametric timed model checker \imitator{}~\cite{Andre21}.
The procedure may not terminate but is exact (sound and complete) if it does; if the synthesis is interrupted before termination (\eg{} using a timeout), then the result is an over-approximation of the set of conditions for which the system is non-interfering.
The method is experimented on a model of the Fischer's mutual exclusion protocol extended from~\cite{BT03}.

\begin{figure}[tb]
		\centering
	 \footnotesize

	\begin{tikzpicture}[pta, scale=1, xscale=1, yscale=1]

		\node[location, initial] at (0, 0) (l0) {$\loc_0$};

		\node[location] at (3, 2) (l1) {$\loc_1$};

		\node[location] at (3, 0) (l2) {$\loc_2$};

		\path (l0) edge[loop above] node[above,align=center]{$\clockx = 2$ \\ $\acth$ \\ $ \clockx \assign 0 $} (l0);
		\path (l0) edge[loop below] node[below,align=center]{$\clockx = 3$ \\ $\acth$ \\ $ \clockx \assign 0 $} (l0);

		\path (l0) edge node[above,yshift=11,align=center]{$\actli{1}$ \\ $ \clocky = \paramp \land \clockx = 0 $} (l1);
		\path (l0) edge node[above,align=center]{$\actli{1}$ \\ $ \clocky = 5 $} (l2);
		\path (l0) edge[bend right] node[below,align=center]{$\actli{2}$ \\ $ \clocky = 4 \land \clockx = 0 $} (l2);

	\end{tikzpicture}

\caption{Illustrating parametric $n$-non-interference}
\label{figure:example-pn-non-interference}

\end{figure}
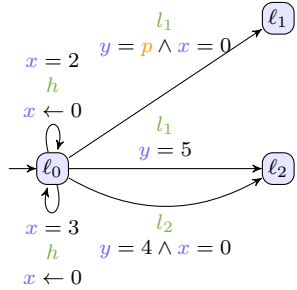
\begin{example}
	First consider again the TA in \cref{figure:example-n-non-interference}.
	Assume that $n$ is now a parameter.
	The method proposed in~\cite{AK20} can show that the TA is non-interferent (\wrt{} locations) iff $n \geq 4$.
	
	Now consider the PTA~$\A$ in \cref{figure:example-pn-non-interference}, which acts as a parametric version of \cref{figure:example-n-non-interference}.
	The PTA is non-interferent iff
	\[
		(\param \in \{ 0,3,8 \})
		\lor
		(n \leq 2 \land \param \in \{ 4,5 \})
		\lor
		(n \leq 3 \land \param \in \{ 2,6,7,9 \})
	\text{.}
	\]
	Observe that this constraint involves both the timing parameter from the model ($\param$) and the frequency ($n$).

\end{example}
\subsection{Non-interference and simulation}\label{ss:noninterference+sim}

In~\cite{GMR07}, Gardey \etal{} propose several definitions of non-interference, related to various notions of simulation.
They consider not only the \emph{verification} problem (``is the system non-interferent?'')\ but also the \emph{control} problem (``synthesize a controller that will restrict the system in order to obtain non-interference'').

They define a notion of timed strong non-deterministic non-interference (timed SNNI) based on timed language equivalence between the automaton with hidden high-level actions ($\hide{\A}{\ActionsH}$) and the automaton with removed high-level actions ($\restrict{\A}{\ActionsL}$).
This notion is similar to $n$-non-interference from~\cite{BDST02} when $n=0$.
As seen previously, this problem is undecidable, which mainly comes from the undecidability of the language inclusion checking in timed automata.

In contrast, three decidable notions are proposed:
\begin{description}
	\item[Timed cosimulation-based SNNI (``timed CSNNI'')] The TA obtained by hiding high-level actions ($\hide{\A}{\ActionsH}$) can be weakly simulated by the original TA~$\A$.
	It is proved that this is equivalent to checking that $\hide{\A}{\ActionsH}$ can be weakly simulated by $\restrict{\A}{\ActionsL}$ and reciprocally.

	\item[Timed bisimulation-based strong non-deterministic non-interference\FinalACMVersion{ (``timed BSNNI'')}] This notion\arXivVersion{ (called ``timed BSNNI'')} extends timed CSNNI to (weak) bisimulation (\ie{} $\hide{\A}{\ActionsH}$ is weakly bisimilar to $\restrict{\A}{\ActionsL}$).
	
	\item[Timed state SNNI (``timed StSNNI'')] The set of reachable locations in $\restrict{\A}{\ActionsL}$ and~$\A$ are identical.
	This definition is equivalent to $n$-non-interference from~\cite{BT03} when $n=0$.
\end{description}

Decidability of the non-interference verification for these three notions comes from the decidability of simulation~\cite{AIPP00}, bisimulation~\cite{LLW95} and location reachability~\cite{AD94} in TAs, respectively.

It is proved in~\cite{GMR07} that timed BSNNI implies timed CSNNI, which itself implies timed SNNI.
\begin{figure}[tb]

	\centering
	 \footnotesize

	\begin{tikzpicture}[pta, scale=1, xscale=1.3, yscale=1]

		\node[location, initial] at (0, 0) (l0) {$\loc_0$};
		\node[location] at (-1, -1) (l1) {$\loc_1$};
		\node[location] at (1, -1) (l2) {$\loc_2$};
		\node[location] at (-1, -2) (l3) {$\loc_3$};
		\node[location] at (+1, -2) (l4) {$\loc_4$};

		\node[location] at (3, 0) (l5) {$\loc_5$};
		\node[location] at (3, -1) (l6) {$\loc_6$};
		\node[location] at (2, -2) (l7) {$\loc_7$};
		\node[location] at (4, -2) (l8) {$\loc_8$};
		
		\path (l0) edge node[]{$\styleact{l_1}$} (l1);
		\path (l0) edge node[]{$\styleact{l_1}$} (l2);
		\path (l1) edge node[align=center]{$\styleclock{x} > 2$\\$\styleact{l_2}$} (l3);
		\path (l2) edge node[]{$\styleact{l_3}$} (l4);

		\path (l0) edge node[]{$\acth{}$} (l5);

		\path (l5) edge node[]{$\styleact{l_1}$} (l6);
		\path (l6) edge node[align=center]{$\styleclock{x} > 2$\\$\styleact{l_2}$} (l7);
		\path (l6) edge node[]{$\styleact{l_3}$} (l8);

		\end{tikzpicture}
	\caption{A TA which is SNNI but not cosimulation-based SNNI (inspired by \cite[Fig.5a]{BCLR15})}
	\label{figure:SNNIvsCSNNI}

\end{figure}
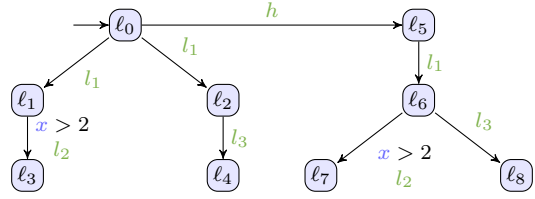
\begin{example}
Assuming $\ActionsH = \{\acth\}$ and $\ActionsL = \{ \actli{1}, \actli{2} \}$, the TA~$\A$ in \cref{figure:SNNIvsCSNNI} is SNNI because the timed language is the same when hiding $\acth{}$ (\ie{} $\hide{\A}{\ActionsH}$) or when pruning $\acth{}$ (\ie{} $\restrict{\A}{\ActionsL}$, that is removing $\acth{}$ and all the subgraph from $\loc_5$).
It is not cosimulation-based SNNI because the automaton obtained when pruning $\acth{}$ (\ie{} $\restrict{\A}{\ActionsL}$) cannot simulate the one when hiding $\acth{}$ (\ie{} $\hide{\A}{\ActionsH}$).
That is because in this case, location $\loc_6$ cannot be simulated by any reachable location.
\end{example}

\begin{figure}[tb]
\centering
\begin{subfigure}{.5\textwidth}
  \centering
  	\begin{tikzpicture}[pta, scale=1, xscale=1.8, yscale=1]

		\node[location, initial] at (0, 0) (l0) {$\loc_0$};
		\node[location] at (0, -1) (l1) {$\loc_1$};
		\node[location] at (1, 0) (l2) {$\loc_2$};
		
		\path (l0) edge node[left,align=center]{$\styleclock{x} \geq 2$\\$\actl{}$} (l1);
		\path (l0) edge node[align=center]{$\acth{}$} (l2);

	\end{tikzpicture}
	 \caption{A TA which is cosimulation-based SNNI but not bisimulation-based SNNI}
	 \label{figure:CNNIvsBSNNI_a}
\end{subfigure}%
\begin{subfigure}{.5\textwidth}
  \centering
	\begin{tikzpicture}[pta, scale=1, xscale=1.8, yscale=1]

		\node[location, initial] at (0, 0) (l0) {$\loc_0$};
		\node[location] at (0, -1) (l1) {$\loc_1$};
		\node[location] at (1, 0) (l2) {$\loc_2$};
		\node[location] at (1, -1) (l3) {$\loc_3$};
		
		\path (l0) edge node[left,align=center]{$\styleclock{x} \geq 2$\\$\actl{}$} (l1);
		\path (l0) edge node[align=center]{$\acth{}$} (l2);
		\path (l2) edge node[align=center]{$\styleclock{x} \geq 2$\\$\actl{}$} (l3);

	\end{tikzpicture}
	  \caption{A TA that is bisimulation-based SNNI}
	  \label{figure:CNNIvsBSNNI_b}
\end{subfigure}
	\caption{Comparing cosimulation-based SNNI with bisimulation-based SNNI (inspired by \cite[Fig.6]{BCLR15})}
	\label{figure:CNNIvsBSNNI}
\end{figure}
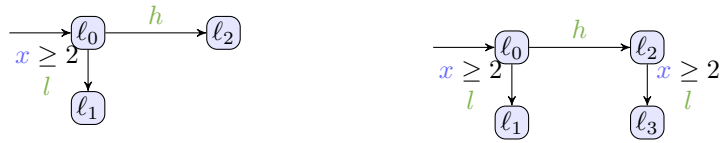

\begin{example}
The automaton in \cref{figure:CNNIvsBSNNI_a} is cosimulation-based SNNI as the automaton obtained by pruning $\acth{}$ (\ie{} $\restrict{\A}{\ActionsL}$) can simulate the one obtained by hiding $\acth{}$ (\ie{} $\hide{\A}{\ActionsL}$)---and reciprocally.
It is not bisimulation-based SNNI since the location $\loc_2$ has no equivalent in the automaton obtained by pruning $\acth{}$ (\ie{} $\restrict{\A}{\ActionsL}$).
On the other hand, the automaton in \cref{figure:CNNIvsBSNNI_b} is bisimulation-based SNNI as the automaton obtained when pruning $\acth{}$ (\ie{} $\restrict{\A}{\ActionsL}$) can simulate the one obtained by hiding $\acth{}$ (\ie{} $\hide{\A}{\ActionsL}$)---and reciprocally.
\end{example}
\paragraph{Control}
The timed non-interference control problem is then addressed in~\cite{GMR07} for those definitions, in the form of an algorithm generating a controller with the assumption that high-level actions are controllable (\ie{} can be enabled or disabled).
It is proved that the timed StSNNI and the timed CSNNI control problems are decidable.

In~\cite{BCLR15}, Benattar \etal{} extend~\cite{GMR07} by proposing a subclass of TAs (called dTA) for which the timed SNNI verification problem becomes decidable.
A model is in the dTA class if the TA obtained from~$\A$ by removing high-level actions (\ie{} $\restrict{\A}{\ActionsL}$) is deterministic.
It is proved that the timed SNNI control problem is decidable on the dTA class, and the synthesized controller is the most permissive one (the one that forbids the least actions);
note that the set of controllable actions is not necessarily~$\ActionsH$, in contrast to~\cite{GMR07}.

Its is possible to synthesize a most permissive controller for timed BSNNI and CSNNI on the dTA class, but not in the general TA class.
The generation of the most permissive controller (for dTAs) is done by solving safety timed games~\cite{CDFLL05}.
\subsection{Non-interference and probabilistic timed automata}\label{ss:noninterference+prob}

A first attempt to adapt timed non-interference to probabilistic timed automata was performed in~\cite{LMT03}.
Lanotte \etal{} prove the decidability of weak bisimulation for probablistic TAs and give a correct algorithm for this problem.
It is then applied to a cryptographic protocol of probabilistic non-repudiation.

Lanotte \etal{} continue this work in~\cite{LMT04,LMT10}, were they propose a framework capturing both timed and probabilistic aspects of non-interference.
Modeling is performed with probabilistic TAs, allowing to analyze information knowledge on probabilistic protocols, where information can be known modulo some probability.
For that purpose, a notion of Probabilistic Timed Non-Interference (PTNI) is defined based on weak bisimulation: this notion of PTNI is decidable.
It is basically the same notion than the timed bisimulation-based SNNI of~\cite{GMR07}, with the addition of the probabilistic behavior.
A notion of Probabilistic Timed Non Deducibility on Composition (PTNDC) is also defined such that it is satisfied on a system where the observational behavior of the isolated system (where high-levels actions are hidden) is equal to the behavior of the system communicating with any high level agent.
An application of the framework is performed on a network device, and PTNDC is checked on the system, allowing to identify covert channels that could not have been detected without capturing both probabilistic and timed aspects.

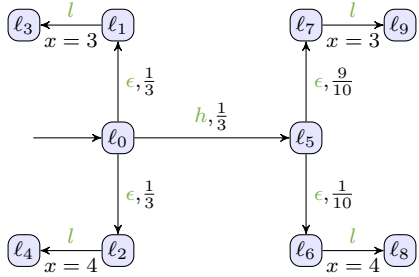
\begin{figure}[tb]

	\centering
	 \footnotesize

	\begin{tikzpicture}[pta, scale=1, xscale=2.5, yscale=1.5]

		\node[location, initial] at (0, 0) (s0) {$\loc_0$};
		\node[location] at (0, 1) (s1) {$\loc_1$};
		\node[location] at (-0.5, 1) (s3) {$\loc_3$};
		\node[location] at (0, -1) (s2) {$\loc_2$};
		\node[location] at (-0.5, -1) (s4) {$\loc_4$};
		\node[location] at (1, 0) (s5) {$\loc_5$};
		\node[location] at (1, -1) (s6) {$\loc_6$};
		\node[location] at (1.5, -1) (s8) {$\loc_8$};
		\node[location] at (1, 1) (s7) {$\loc_7$};
		\node[location] at (1.5, 1) (s9) {$\loc_9$};

		\path (s0) edge node[right]{$\styleact{\actionSilent}$,$\frac{1}{3}$} (s1);
		\path (s1) edge node[above]{$\actl{}$} node[below]{$x=3$} (s3);
		\path (s0) edge node[right]{$\styleact{\actionSilent}$,$\frac{1}{3}$} (s2);
		\path (s2) edge node[above]{$\actl{}$} node[below]{$x=4$} (s4);
		\path (s0) edge node[above]{$\acth{}$,$\frac{1}{3}$} (s5);
		\path (s5) edge node[right]{$\styleact{\actionSilent}$,$\frac{1}{10}$} (s6);
		\path (s6) edge node[above]{$\actl{}$} node[below]{$x=4$} (s8);
		\path (s5) edge node[right]{$\styleact{\actionSilent}$,$\frac{9}{10}$} (s7);
		\path (s7) edge node[above]{$\actl{}$} node[below]{$x=3$} (s9);

	\end{tikzpicture}
	\caption{Probabilistic timed non-interference \cite[Fig.6]{LMT10}}
	\label{figure:PTNI}

\end{figure}
\begin{example}
\cref{figure:PTNI}, taken from~\cite[Fig.6]{LMT10}, presents a system where probabilistic timed non-interference is violated.
If the high-level action $\acth{}$ is not taken, one can observe that the timed language is either $\{(\actl{},3)\}$ or $\{(\actl{},4)\}$, with a probability of $\frac{1}{2}$ in both cases.
However, if high-level action $\acth{}$ is taken, the probability of observing $(\actl{},3)$ is $\frac{9}{10}$, and the probability of observing $(\actl{}, 4)$ is $\frac{1}{10}$.
Thus, $\acth{}$ interferes with the probability of observing $\actl{}$.
Note that the system becomes non-interferent if either probabilities or time are not considered. %
\end{example}
\subsection{Timed opacity}\label{ss:timed-opacity}

The notion of timed opacity is fairly close to timed non-interference.

Opacity was initially introduced in~\cite{Mazare04,BKMR08} to model information leaks from a system to an attacker; that is, it expresses the power of the attacker to deduce some secret information based on some publicly observable behaviors.
In~\cite{Cassez09}, Cassez extends this notion to \emph{timed} opacity.
If an attacker observing a subset of the actions cannot deduce whether a given sequence of actions has been performed, then the system is opaque. %
Here, the attacker can also observe time, hence untimed systems that are opaque may not be so once time is added.
The paper proves that, with such a definition, opacity is undecidable, even for the restricted class of event-recording automata (ERAs)~\cite{AFH99}, a strict subclass of TAs (\ie{} strictly less expressive than TAs).

In~\cite{WZ17}, Wang \etal{} define a notion of initial-state opacity that is proved decidable on real-time automata~\cite{Dima01} (RTAs).
The RTA formalism is a strict subclass of TA with a single clock which is reset at each transition, meaning the only timed information expressed by this formalism is the time elapsed in the current location; most properties are decidable, even complement and language inclusion (undecidable for the full class of TAs).
In this context, an intruder can observe the elapsed time on a given subset of transitions, and the system is opaque if an intruder cannot determine whether the system starts or not from a given secret state.
The method described in this paper consists in constructing two RTAs accepting the projection of languages, from secret initial states for the first one, and from non-secret ones for the other.
The authors then use a notion of trace equivalence between an RTA and a finite-state automaton in order to translate the two RTAs into their respective trace equivalent (untimed) finite-state automata.
If the intersection of those two automata is empty, the system is initial-state opaque.

In~\cite{WZA18} the authors extend this work to timed language-opacity (\ie{} the opacity on timed language studied in~\cite{Cassez09}).
The positive decidability properties of RTAs allow to check the emptiness of the inclusion between the language accepted by the system and the language where secret words are not allowed.
Therefore, timed language-opacity is decidable on the RTA subclass.
A tool has been developed to check language-opacity on RTA models.

A recent work by Ammar \etal{} \cite{AEYM21} addresses the timed opacity problem in a similar fashion to~\cite{Cassez09}: an intruder has access to a subset of actions, along with timed information.
The originality of~\cite{AEYM21} is to consider a \emph{time-bounded} framework.
As in~\cite{Cassez09}, a secret location is timed opaque if the intruder cannot infer from the observation of any execution that the system has reached this particular location.
Ammar \etal{} propose two variants of this definition of opacity.
The first one, called \emph{timed bounded opacity} implies that the location is opaque up to a given time duration, and is decidable for non-Zeno TA (\ie{} TA where it is impossible for an infinity of actions to occur in finite time).
The second one, \emph{$\delta$-duration bounded opacity} implies that the location must remain secret for at least $\delta$ time units after it is reached.
To solve the $\delta$-duration bounded opacity, one needs to construct two TAs, with final secret and non-secret locations respectively, then apply a reduction such that, for any accepted word $\word$ of timed length $|\word|$, all prefixes $\word'$ of that word have a timed length $|\word'| \geq |\word| - \delta$.
The problem can then be solved by the timed bounded inclusion checking, which is decidable for timed automata~\cite{ORW09}; 
this is in contrast with the (potentially unbounded) language inclusion checking, known to be undecidable~\cite{AD94}.
This time-bounded setting is the crux that explains the difference of decidability between~\cite{Cassez09} and~\cite{AEYM21}.
A case study on a cloud service model illustrates the method by determining the $\delta$ value up to which the system is $\delta$-duration bounded opaque.
The verification is run using \SpaceEx{}~\cite{FLDCRLRGDM11}, a tool taking as input hybrid automata~\cite{Henzinger96}, a formalism of which TAs are a subclass.

\begin{figure}[tb]

	\centering
	 \footnotesize
	 
	\begin{subfigure}{.32\textwidth}
		\centering
		\begin{tikzpicture}[pta, scale=1, xscale=1.2, yscale=1.5]

			\node[location, initial] at (0, 0) (s0) {$\loc_0$};
			\node[location,private] at (1.2, 0.3) (s1) {$\loc_1$};
			\node[location] at (2.4, 0.3) (s2) {$\loc_2$};
			\node[location] at (1.2, -0.3) (s3) {$\loc_3$};
			\node[location] at (2.4, -0.3) (s4) {$\loc_4$};

			\path (s0) edge[bend left] node[above]{$\styleact{a}$, $\clockx \leq 2$} (s1);
			\path (s1) edge node[above]{$\styleact{b}$, $\clockx \leq 7$} (s2);
			\path (s0) edge[bend right] node[below]{$\styleact{a}$, $\clockx \leq 2$} (s3);
			\path (s3) edge node[below]{$\styleact{b}$, $\clockx \leq 6$} (s4);

		\end{tikzpicture}
		\caption{Illustrating \cite{AEYM21}}
		\label{figure:example-AEYM21}
	\end{subfigure}
	\begin{subfigure}{.32\textwidth}
		\centering
		\begin{tikzpicture}[pta, scale=1, xscale=1.2, yscale=1.5]

			\node[location, initial] at (0, 0) (s0) {$\loc_0$};

			\node[location, private] at (2, 0) (s2) {$\loc_S$};

			\node[location, final] at (1, -.5) (s1) {$\loc_T$};

			\node[invariant, above=of s0] {$\clockx \leq 3$};
			\node[invariant, above=of s2] {$\clockx \leq 3$};

			\path (s0) edge node[align=center]{$\clockx \geq \styleparam{\param_1}$} (s2);
			\path (s0) edge[bend right] node[left, align=center]{$\clockx \geq \styleparam{\param_2}$} (s1);
			\path (s2) edge[bend left] node[align=center]{} (s1);

		\end{tikzpicture}
		\caption{Illustrating \cite{ALMS22}}
		\label{figure:example-ALMS22}
	\end{subfigure}
	\begin{subfigure}{.32\textwidth}
		\centering
		\begin{tikzpicture}[pta, scale=1, xscale=1, yscale=1.5]
	
			\node[location, initial] at (0, 0) (s0) {$\loc_0$};
			\node[location] at (2, 0) (s1) {$\loc_1$};
			\node[location,private] at (3.5, 0) (s2) {$\loc_2$};

			\path (s0) edge[bend right] node[below]{$\styledisc{h} > 0 \land \clockx > 1$} (s1);
			\path (s0) edge[bend left] node[above]{$\styledisc{h} \leq 0 \land \clockx \leq 1$} (s1);
			\path (s1) edge node[above]{$\clockx > 1$} (s2);
	
		\end{tikzpicture}
		\caption{Illustrating \cite{VNN18}}
		\label{figure:example-VNN18}
	\end{subfigure}

	\caption{Illustrating various notions of non-interference}
\end{figure}
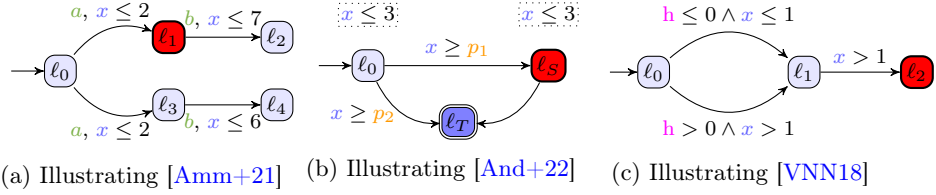
\begin{example}
	Here, we illustrate the $\delta$-duration bounded opacity from~\cite{AEYM21}.
	Consider the TA in \cref{figure:example-AEYM21} (taken from~\cite[Fig.2a]{AEYM21}) where $\styleact{a}$ and $\styleact{b}$ are actions of the model and $\loc_1$ is a secret location.
	If we define $\delta = 4$, $\loc_1$ is $\delta$-duration bounded opaque: indeed, for every timed word passing through~$\loc_1$ and ending $\delta$ time units after reaching~$\loc_1$, there exists an equivalent timed word not passing through~$\loc_1$.
	However, if we consider $\delta = 5$ or higher, $\loc_1$ is not $\delta$-duration bounded opaque anymore, as there now exist timed words passing through~$\loc_1$ and ending $\delta$ time units after reaching~$\loc_1$ such that there exists no equivalent timed word not passing through~$\loc_1$.
	This is for example the case of the timed word ``$(\styleact{a},2) (\styleact{b},7)$''.
\end{example}

In~\cite{ALMS22}, André \etal{} propose a method to check timed opacity, where the attacker can only observe the ``execution time'', \ie{} the runs durations from the initial location to a given target location~$\loc_T$.
This differs from~\cite{Cassez09}, where the attacker was able to observe some actions with their timestamps.
More precisely, the definition of~\cite{ALMS22} states that a system is opaque with regard to a secret location~$\loc_S$ on the way to a target location $\loc_T$ if, for some ``executions times'', $\loc_T$ can be reached both with or without passing through~$\loc_S$.
Two problems are considered:
\begin{oneenumerate}%
	\item the timed opacity problem, that computes the set of execution times (from the initial location to the target location) for which the system is opaque, \ie{} it is not possible to deduce whether the secret location~$\loc_S$ was visited; and
	\item the full timed opacity problem, that checks whether the system is opaque for \emph{all} execution times.
\end{oneenumerate}%
These two notions of opacity are proved to be decidable on (non-parametric) TAs.
That is to say, the attacker model of~\cite{ALMS22} is weaker than that of~\cite{Cassez09}: while an attacker in~\cite{Cassez09} can read some actions with their timestamp, the attacker in~\cite{ALMS22} has access only to the system execution time, which can be seen as a single action with its timestamp (\ie{} the action denoting system completion).

\paragraph{Parametric timed opacity}
The notion of opacity considered in~\cite{ALMS22} is extended to parametric TAs in the same work.
First, the \emph{timed opacity emptiness} problem is considered: does there exist a parameter valuation and an execution time for which the system is opaque?
This problem is shown to be undecidable for general PTAs, but decidable on a sub-class of parametric timed automata called L/U-PTAs~\cite{HRSV02}. %
Second, the \emph{full timed opacity emptiness} problem is considered: does there exist a parameter valuation for which the system is opaque for all execution times?
This second problem is undecidable both for general PTAs and for the aforementioned L/U-PTAs.
Third, the timed opacity \emph{synthesis} problem is addressed, \ie{} the computation of both execution times and parameter valuations guaranteeing opacity.
Despite undecidability of the emptiness problem, a semi-algorithm is provided (\ie{} that may not terminate, but is correct if it does).
The method is demonstrated on a set of benchmarks, including standard PTAs benchmarks~\cite{AMP21}, as well as Java programs from the \href{https://github.com/Apogee-Research/STAC/}{DARPA Space/Time Analysis for Cybersecurity} (STAC) library, manually translated into PTAs.
Experiments are run using \imitator{}, and allow to synthesize the timing parameter valuations for which this definition of opacity holds on the system.
The full time opacity synthesis problem is however left out.
An artifact allowing reproducibility is available online.

\begin{example}
	Consider the PTA in \cref{figure:example-ALMS22} (taken from~\cite{ALMS22} and inspired by~\cite[Fig.2]{GMR07}).
	Consider the parameter valuation (\ie{} replacing a parameter with a concrete valuation) $\pval$ such that $\pval(\param_1) = 1$ and $\pval(\param_2) = 2$.
	The resulting TA is time-opaque \wrt{} $\loc_S$ on the way to~$\loc_T$ for execution times $[2,3]$ because, for any duration $d$ with $d \in [2,3]$, there exists both a run passing by $\loc_S$ and a run \emph{not} passing by $\loc_S$ and reaching $\loc_T$ in $d$ time units.
	This is not the case of all execution times: for example, for $d = 1.5$ there exists a run of duration~1.5 passing by~$\loc_S$ and reaching~$\loc_T$, but no such run \emph{not} passing by~$\loc_S$.

	In the parametric setting, the method given in~\cite{ALMS22} allows to synthesize parameter valuations for which the TA is (full) time-opaque.
	Here, for valuations~$\pval$ such that $\pval(\param_1) = \pval(\param_2)$, the system is %
	time-opaque for all execution times.
\end{example}
\subsection{Non-interference and information flow}\label{ss:noninterference+flow}

In~\cite{NNV17}, Vasilikos \etal{} describe a \emph{timed command language} adapted from Dijkstra's guarded command language~\cite{D75}, and show how specifications of this language can be translated to a non-deterministic subclass of TAs.
A property of non-interference can be verified on the type system~\cite{Cardelli96} of the specified program, ensuring that there is no flow of information from a high-level action to a low-level one.
The approach is demonstrated on a small voting protocol.
This method is limited to the subclass of automata than can be described in the command language, and its scalability is not discussed.
\uppaal{} is used as the underlying model checker, but no automated translation seems to be proposed.

In~\cite{VNN18}, Vasilikos \etal{} develop a method to check soundness of timed systems allowing intentional information leaks, such as smart power grids~\cite{GSKEBCH11}.
The system is modeled with TAs where locations are either strongly or weakly observable in regard to an attacker.
In that sense, \cite{VNN18} addresses a broader class than~\cite{NNV17}, that was limited to TAs that can be described by the timed command language.
The modeling of~\cite{VNN18} describes where it is allowed to leak information and where it is not (weakly observable locations being allowed to bypass the security policy).
The information flow is not described through actions but through variables (low-level ones, that are visible, and high-level ones, that must remain hidden).
An algorithm is given that checks that no information flow leaks on strongly observable locations.
As in~\cite{NNV17}, no implementation seems to be available.

 \begin{example}
 	Let us briefly illustrate some of the notions from~\cite{VNN18}.
 	Consider the TA in \cref{figure:example-VNN18}.
 	We assume that $\styledisc{h}$ is a high-level variable, $\clockx$ is a clock, and $\loc_2$ is strongly observable.
 	If $\loc_1$ was strongly observable,
		non-interference would not hold, as observing the clock $\clockx$ allows to deduce whether or not $\styledisc{h}$ has a negative value.
 	However, the guard from $\loc_1$ to $\loc_2$ erases this information in such a way that the system respects non-interference.
 \end{example}

In~\cite{VNNK19}, Vasilikos \etal{} study the impact of \emph{clock} granularity on timing channel attacks.
Timing side-channels attacks are closely related to non-interference, as an attacker is able to deduce internal information about a system by accessing some publicly available data---notably time.
Here, by ``clock granularity'', we mean \emph{clock} as in \emph{processor clock} that ticks periodically (different from the abstract concept of ``timed automata clocks'').
The authors in~\cite{VNNK19} notably propose a modeling based on timed automata with stochastic conditions, and they provide sufficient conditions ``for when one can achieve better security by increasing the grain of the clock''.
The authors describe stochastic systems with the help of probabilistic TAs, and provide an algorithm for obtaining, given an attack scenario on a deterministic system, the timing channels along with their respective probability of occurrence.
Models in TAs are given for attacks techniques from the literature (the one-pad technique, the clock-edge technique and the co-prime technique)~\cite{SMGM17}.
No implementation is provided, but a manual proof on the case study is considered.

In~\cite{GSB18}, Gerking \etal{} present a method reducing the verification of timed non-interference to a refinement check proposed in~\cite{HBDS15}, itself reducing to checking ``testing automata'', \ie{} reducing to reachability~\cite{ABBL03}.
Their notion of timed non-interference is equivalent to the SNNI from~\cite{GMR07}, where it is expressed as a bisimulation of the low-level behavior between the original automaton and the automaton with high-level actions disabled.
For the refinement check, the automata of the latter is modified by adding a dedicated error location that is reachable if and only if timed non-interference is violated.
If the error location is reachable on the parallel composition of the two automata, the system is interferent.
This technique, however, requires that the modified automaton is deterministic, resulting in the same expressive power as the work of Benattar \etal{} \cite{BCLR15}, where the SNNI verification problem is proved to be decidable on the dTA subclass (the automaton obtained by removing high-level actions must be deterministic).
The method is experimented using \uppaal{} on a model of a cyber-manufacturing system, where it is able to detect a potential timing channel attack, and propose a mitigation by adding some temporal constraints on the system.
The approach of non-interference based on reachable locations~\cite{BT03} is also used as a comparison, and is not able to detect the attack.

Finally note that~\cite{DXS18} proposes a formal method based approach \emph{not} using TAs, and therefore not detailed in this survey, but not entirely unrelated either: their approach identifies CTL formulas that express paths that could lead to timing attacks.
Bounded model checking techniques can be applied, and \uppaal{} (that takes TAs as input formalism) is cited as one of the possible target model checkers.

\subsection{Discussion}\label{ss:noninterference:discussion}

\cref{table:summary-non-interference} summarizes the results of each paper surveyed in this section.
The second column indicates the kind of TNI (timed non-interference) property studied by the paper, while the third column indicates the class of models on which the work applies (dTA is a subclass such that the automaton obtained by removing high-level actions is deterministic~\cite{BCLR15}).
``Finite-state automata'' indicates that the work applies not only to TAs but to any kind of automata-based formalism.
The last three columns indicate whether the approach supports controller synthesis, whether the property studied in the paper is decidable, and whether the work is supported by an automated model checking tool, respectively.

\begin{table}[tb]
\caption{Comparison of works on timed non-interference.}%
\label{table:summary-non-interference}%
{\centering
\scriptsize
\setlength{\tabcolsep}{2pt} %
\begin{tabular}{ |l|c|c|c|c|c| }
\hline
	\rowHeader{} Work & TNI property & Class of models & Control & Decidable & Tool\FinalACMVersion{ support}\\
\hline
Lanotte \etal{} \cite{LMT03} & Weak bisimulation  & Probabilistic TA & \cellNo{} & \cellYes{} & \cellNo{} \\ 
\hline
Gardey \etal{} \cite{GMR07} & Weak bisimulation & TA & \cellYes{} & \cellNo{} & \cellNo{} \\ 
\hline
Benattar \etal{} \cite{BCLR15} & Weak bisimulation & TA subclass (dTA)  & \cellYes{} & \cellNo{} & \cellNo{} \\ 
\hline
\hline
Barbuti \etal{} \cite{BDST02} & TNI on language & TA  & \cellNo{} & \cellNo{} & \cellNo{} \\
\hline
Giacobazzi\FinalACMVersion{ and Mastroeni}\arXivVersion{ \etal{}} \cite{GM05} & TNI on language & Finite-state automata  & \cellNo{} & \cellNo{} & \cellNo{} \\
\hline
Lanotte \etal{} \cite{LMT04,LMT10} & subset of TNI on Language & Probabilistic TA & \cellNo{} & \cellYes{} & \cellNo{} \\ 
\hline
Gardey \etal{} \cite{GMR07} & Cosimulation TNI & TA & \cellYes{} & \cellYes{} & \cellNo{} \\ 
\hline
Benattar \etal{} \cite{BCLR15} & TNI on language & TA subclass (dTA)  & \cellYes{} & \cellYes{} & \cellNo{} \\ 
\hline
Gerking \etal{} \cite{GSB18} & TNI on language & TA subclass (dTA) & \cellNo{} & \cellYes{} & \uppaal{} \\
\hline
\hline
Barbuti and Tesei\cite{BT03} & TNI on reachable states & TA  & \cellNo{} & \cellYes{} & \cellNo{} \\
\hline
Gardey \etal{} \cite{GMR07} & TNI on reachable states & TA & \cellYes{} & \cellYes{} & \cellNo{} \\ 
\hline
André and Kryukov \cite{AK20} & TNI on reachable states & Parametric TA  & \cellNo{} & \cellYes{} & \imitator{} \\
\hline
\hline
Cassez \cite{Cassez09} & Timed opacity on language & ERA  & \cellNo{} & \cellNo{} & \cellNo{}\\
\hline
Wang \etal{} \cite{WZA18} & Timed opacity on language & TA subclass (RTA)  & \cellNo{} & \cellYes{} & \cellYes{}\\
\hline
Ammar \etal{} \cite{AEYM21} & Timed opacity on language (bounded) & non-zeno TA  & \cellNo{} & \cellYes{} & \SpaceEx{}\\
\hline
\multirow{5}{*}{André \etal{} \cite{ALMS22}} & Timed opacity  (execution times) & TA & \cellNo{} & \cellYes{} & \imitator{} \\
\cline{2-6}
 & Full timed opacity (execution times) & TA & \cellNo{} & \cellYes{} & \cellNo{} \\
\cline{2-6}
 & Timed opacity (execution times) & Parametric TA & \cellNo{} & \cellNo{} & \imitator{} \\
\cline{2-6}
 & Timed opacity emptiness & Parametric TA subclass & \cellNo{} & \cellYes{} & \imitator{} \\
\cline{2-6}
 & Full timed opacity (execution times) & Parametric TA subclass & \cellNo{} & \cellNo{} & \cellNo{} \\
\hline
\hline
Nielson \etal{} \cite{NNV17} & TNI on variables & TA subclass & \cellNo{} & \cellYes{} & \cellNo{} \\ 
\hline
Vasilikos \etal{} \cite{VNN18}  & TNI on variables & TA & \cellNo{} & \cellYes{} & \cellNo{} \\ 
\hline
Vasilikos \etal{} \cite{VNNK19}  & Clock granularity & Probabilistic TA & \cellNo{} & \cellYes{} & \cellNo{} \\ 
\hline
\end{tabular}

}
\end{table}

The framework of~\cite{BCLR15} seems to be one of the most important works so far as it allows not only to check non-interference on reachable states on TA and non-interference on accepted language for the dTA subclass, but also to generate a controller ensuring such properties on the system.

\cite{WZA18} allows to check language-opacity on real-time automata, made possible by the strong syntactic restrictions on this subclass of TAs, while~\cite{AEYM21} allows to check location-based language opacity (over bounded time) on non-Zeno timed automata.
\cite{VNN18} allows to check non-interference on variable flow on TAs.
\cite{LMT10} allows to check a subset of non-interference on accepted language (non-deducibility on composition) on probabilistic~sTA.
Concerning parametric~TAs, \cite{AK20} addresses the synthesis of timing parameters for which $n$-non-interference holds, while~\cite{ALMS22} allows to check a rather weak notion of timed opacity on TAs and on a subset of parametric~TAs.
Together, those works combine all that is currently expressible on timed non-interference.

A major outcome from \cref{table:summary-non-interference} is that only few works benefit from an actual implementation.
This is not only disappointing, but also surprising, as the TA community came out with several efficient reasonably scalable model checkers---the leading one being \uppaal{}, already used in a large number of industrial contexts.
Therefore, a gap seems to be still present between the \emph{theory} of non-interference in the context of TAs (``which problem \emph{can} be answered?''), and the actual \emph{automated verification} of non-interference properties of timed systems using TA-based techniques.

Finally note that only one work uses \uppaal{}, while others use tools applied to \emph{extensions} of TAs: \SpaceEx{} (that takes as input \emph{hybrid} automata) and \imitator{} (that takes as input \emph{parametric} timed automata).
This may come from the fact that opacity problems are in general harder than reachability, and \uppaal{} is a model checker mainly targeting reachability properties.

\section{Attack trees}\label{section:attacktrees}

Attack trees~\cite{KPS14} are a formalism inspired by fault trees~\cite{VGRH81,RS15}, that represent the possible behaviors of an attacker.
An attack tree takes the form of a \emph{tree}, where the \emph{root} represents the goal of the attack.
Each node, starting by the root, is refined with logical gates (such as ``AND''/``OR'') into sub-goals, until basic actions are reached, forming the \emph{leaves} of the tree.

\begin{figure}[tb]

	\includegraphics[width=\textwidth]{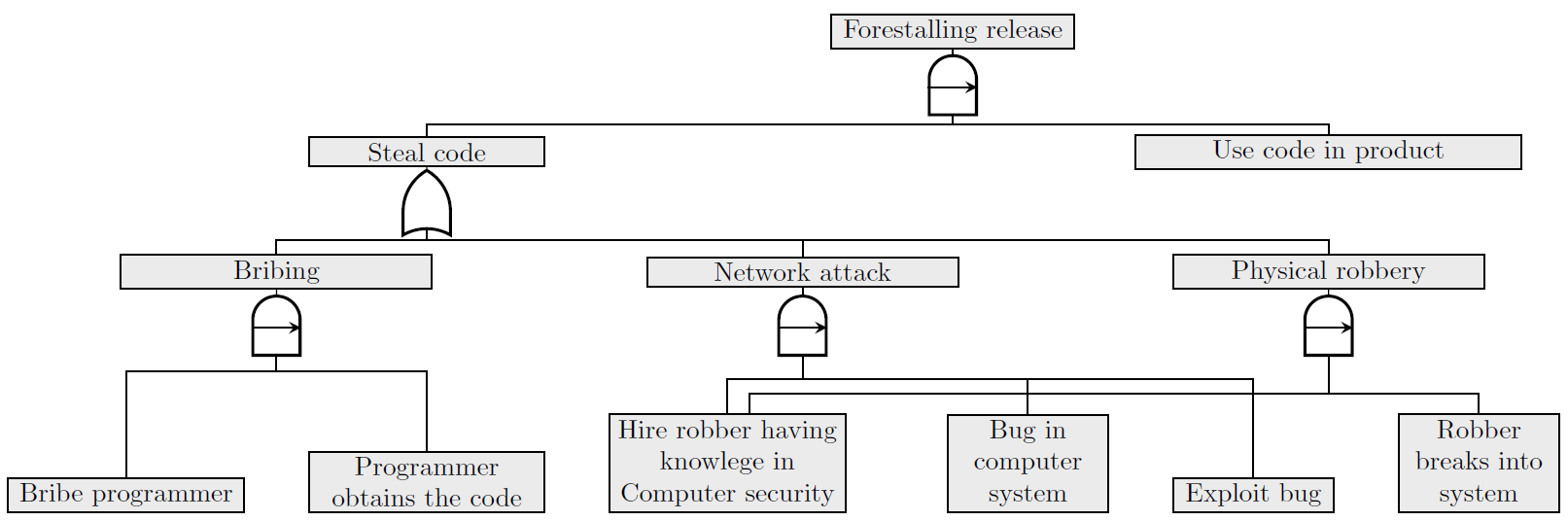}

	\caption{Attack tree modeling the forestalling of a software \cite[Fig.1]{KRS15}}
	\label{figure:ExampleAT}

\end{figure}
\begin{example}
\cref{figure:ExampleAT}, taken from~\cite[Fig.1]{KRS15}, depicts an attack tree modeling how one could get access to a software code.
Stealing the code (OR gate) can be done in three different ways: Bribing, a Network Attack or a Physical robbery.
Each of those possibilities is modeled by a SAND (``sequential AND'') gate, meaning that each subgoal must be executed from left to right.
For instance, successfully Bribing implies first that a programmer is bribed, and then that the bribed programmer obtains the code.
\end{example}

Since~2015, a line of works started to use the theory (and tools) of timed automata in order to equip attack trees with timing information, and model and verify them using TA software.
In the following, we review several works encoding attack trees (and their extensions) into timed automata (and their extensions).

\subsection{Attack-fault trees}

The first use of TAs as a target formalism for attack trees is described in~\cite{KRS15}, where Kumar \etal{} extend attack trees with cost values for several resources, such as time.
Their framework allows to answer cost related security properties.
For example, it is possible to know which attack path minimizes some resources, or what are the paths that are more harmful to the system.
Pareto-optimal curves are used when several resources are to be optimized.
A translation from the extended attack trees to priced timed automata~\cite{BLR04} is described.
A timed automaton is constructed for each node of the attack tree.
The parallel composition of these TAs generates a priced timed automaton encoding the attack tree.
The paper then provides guidelines on how to verify the model against Weighted CTL with \uppaal{}-CORA, an extension of \uppaal{} for priced TAs.

This first work is then extended to attack-fault trees (AFTs) in~\cite{KS17}.
Logical gates of both fault trees and attack trees are allowed in this formalism, thus increasing the expressiveness compared to standard attack trees.
AFTs are translated to probabilistic TAs~\cite{B03} in a similar way to that of~\cite{KRS15}, and checking is performed with \uppaal{}-SMC~\cite{DLLMP15}.
A case study of an oil pipeline from~\cite{KBCHP14} is analyzed.

In~\cite{VGMH20}, Valluripally \etal{} describe a framework for performing model checking on Virtual Reality Learning Environment (VRLE)~\cite{PCYZS06}, which is a part of the Internet of Things.
The framework consists in modeling attack trees and translating them to probabilistic timed automata.
Unlike~\cite{KRS15}, each leaf node is transformed into an automaton that models the corresponding attack scenario.
Model checking can be performed with \uppaal{}-SMC to determine if the probability of disruption of the system is higher than a set threshold.
If it is for some attack scenario, modification of the system's design can be proposed, resulting in a novel attack tree, and the process can continue until the system is considered safe.
The method is illustrated on iSocial~\cite{ZSHNCH18} a VRLE for youth with autism spectrum disorder, composed of VR headset devices, handheld controllers and base stations.
The probability of loss of information or privacy leakage can be studied either for each of the attacks scenarios present on the tree (by checking with the appropriate automaton) or by combining them.
This allows to observe that combining denial of service and unauthorized access always leads to a loss of information (a probability of~1), while combining unauthorized access and user location always leads to a privacy leakage.
The implementation of some design principles (mainly consisting in limiting authorizations to strict functionality) results in a significant reduction in both loss of information and privacy leakage.

Finally, in~\cite{Ali21}, Ali translates a so-called extension of attack trees named ``simplified timed attack trees'' (STAT) into a network of weighted timed automata.
\uppaal{} is the underlying model checker.
No automated translation tool seems to be available, and the novelty compared to the aforementioned works (notably~\cite{KRS15}) remains unclear.

\paragraph{Testing IoT systems using attack trees}

In~\cite{KA19}, Krichen and Alroobaea propose a framework for formal testing IoT systems in a realistic environments, using attack trees.
It borrows the modeling and translation to priced timed automata from~\cite{KRS15}.
The resulting network of automata is used as an input for generating abstract test cases.
Those test scenarios are then translated to the TTCN-3 (Testing and Test Control Notation version 3)~\cite{TP12} specification language for test execution.
The tests are run on a cloud testing architecture that will output verdicts according to the TTCN-3 language.
\paragraph{Adding parameters}

In~\cite{ALRS21}, André \etal{} further extend~\cite{KS17} by devising a method synthesizing the set of all timing parameter and cost parameter valuations leading to a successful attack.
To do so, the AFTs are modeled with parametric weighted timed automata which is an extension of parametric timed automata where transitions are associated with possibly parametric (discrete) costs.
This formalism allows multiple weights, which can be incremented by constants or variables values.
It does not, however, consider stochastic information.
The compiler ATTop~\cite{RYHBRS18} provides an automated translation from AFTs to parametric weighted timed automata.
Similarly to previous approaches, each node of the tree is modeled using a parametric weighted timed automaton.
The resulting model can be checked with \imitator{}.
The method is experimented on two case studies.
\subsection{Attack-defense trees}

Meanwhile, Gadyatskaya \emph{et al.}~\cite{GHLLOP16} extend the attack-defense tree formalism~\cite{KMRS10,KMRS14} (\ie{} attack trees including defender's actions) with a timed semantics.
Stochastic behavior and cost of actions are also added to the semantics.
A translation algorithm that generates a network of probabilistic TAs is provided.
Unlike Kumar \etal{}~\cite{KRS15}, this work actually models the system using the classical approach where each participant is modeled by an automaton.
The attack-defense tree goals are translated into Boolean formulas, the attacker's goal being to reach a state where the Boolean formula encoding the attack-defense tree is true.
The probabilistic extension of timed automata is used both to answer queries on the probability of some event, and to model a cost on the actions taken by the attacker.
Model checking is done with \uppaal{}-Stratego~\cite{DJLMT15} and can answer queries on cost or probabilities of an attack, along with timed constraints.
The method is illustrated on an example, which allows to deduce expecting cost of attack.

This work is then extended by Hansen \etal{} in~\cite{HJLLP17}.
The cost of attacks is modeled as a function of time, depicting a more realistic system.
Also, the attacker is now parametric in a sense that a range of valuations will be checked for parameters defining the probability of non-deterministic behaviors.
Analysis of variance (a method allowing to compare the effect of discrete factors on continuous variables) is proposed to identify the parameter valuations that minimize the cost of an attack.
This time, model checking is performed with \uppaal{}-SMC.

In~\cite{HKKS16}, Hermanns \etal{} propose a formalism that is more expressive than attack-defense tree: attack-defense diagrams (ADDs).
Those allow cycles, capturing the dynamic aspects of attack scenarios.
A node in an ADD represents not only the actions of the attacker/defender, but also independent timed events.
The attacker's goal is to reach a sink node (without outgoing edges) that represents the success of the attack.
Likewise, the defender's goal is to reach another sink node that represents the failure of the attack.
Each node of the resulting graph is translated into a probabilistic timed automaton that models either a timed event, or a logical gate corresponding to an action.
Model checking is performed on the parallel composition of the resulting automata using the stochastic model checking tool Modest~\cite{BDHK06}.
It allows to check properties expressing cost, time or probability of success, and the process is illustrated on an \emph{ad-hoc} case study.

In~\cite{ABPPSS20}, Arias \etal{} adapt the translation from~\cite{ALRS21} to attack-defense trees, aiming in particular at modeling coalitions of agents.
Attack-defense trees are extended with agents and new gates (such as the SCAND---for sequential AND gate with attacks and defenses as children).
The semantics is given in terms of extended asynchronous multi-agent systems (EAMAS), that extend AMAS~\cite{JPSDM20} with attributes in local transitions.
Attack-defense trees are translated into the EAMAS formalism by turning each node in the tree as an agent.
A dedicated tool adt2amas\footnote{\url{https://lipn.univ-paris13.fr/adt2amas/}} takes as input an attack-defense tree, and automatically generates the corresponding EAMAS model.
The tool then translates that EAMAS model into modgraph or 
into a network of parametric timed automata in the \imitator{} format.
Case studies illustrate how the method can answer questions regarding cost and time of attacks.

Petrucci \etal{} extend in \cite{PKPS19} this latter work with a reduction technique dedicated to trees, allowing for a faster checking process.
It consists in using the layered structure present in a tree, in particular an attack tree, to ignore some of the interleavings when exploring the tree.
This latter work was also integrated into the adt2amas tool (with option \texttt{--layered-reductions}).

\subsection{Discussion}

\cref{table:summary-attack-trees} summarizes the results of each paper presented in this section.
Columns 3,4 and~5 (from left to right) respectively indicate whether the tree formalism supports %
	probabilities, cost and/or timing parameters, and agents.
Note that the work from~\cite{HKKS16} not only allows modeling defense aspects, but also introduces cycles and new logical gates to the formalism.
The sixth column indicates the class of models used as a target formalism, which relates to the level of expressiveness of the properties studied.
The right-most column gives the tools used to perform automated verification, if any.

\begin{table}[tb]
\caption{Comparison of works using TA as target formalism for attack trees.}%
\label{table:summary-attack-trees}
{\centering
\scriptsize
\setlength{\tabcolsep}{2pt} %
\begin{tabular}{ |l|c|c|c|c|c|c| }
\hline
	\rowHeader{}
	Work & Input formalism & Prob\FinalACMVersion{abilities} & Param\FinalACMVersion{eters} & Agents & Target formalism & Tool support \\
\hline
Kumar \etal{} \cite{KRS15} & Attack-fault trees & \cellNo{} & \cellNo{} & \cellNo{} & Priced TA & \uppaal{}-CORA \\
\hline
Kumar and Stoelinga \cite{KS17}  &  Attack-fault trees & \cellYes{} & \cellNo{} & \cellNo{}
	& Probabilistic TA & \uppaal{}-SMC \\
\hline
Valluripally \etal{} \cite{VGMH20} & Attack-fault trees & \cellYes{} & \cellNo{} & \cellNo{}
	& Stochastic TA %
& \uppaal{}-SMC \\ 
\hline
André \etal{} \cite{ALRS21} & Attack-fault trees  & \cellNo{} & \cellYes{} & \cellNo{}
	& Parametric weighted TA %
& \imitator{} \\ 
\hline
Ali \cite{Ali21} & Attack-fault trees & \cellNo{} & \cellNo{} & \cellNo{} & Weighted TA & \uppaal{} \\
\hline
Gadyatskaya \etal{} \cite{GHLLOP16} & Attack-defense trees & \cellYes{} & \cellNo{} & \cellNo{}
	& Probabilistic TA & \uppaal{}-Stratego \\ 
\hline
Hansen \etal{} \cite{HJLLP17}  & Attack-defense trees & \cellYes{} & \cellNo{} & \cellNo{}
	& Probabilistic TA
	& \uppaal{}-SMC \\
\hline
Petrucci \etal{} \cite{ABPPSS20,PKPS19} & Attack-defense trees & \cellNo{} & \cellYes{} & \cellYes{}
	& Parametric TA & \uppaal{} / \imitator{} \\
\hline
Hermanns \etal{} \cite{HKKS16} & Attack-defense diagrams & \cellYes{} & \cellNo{} & \cellNo{}
	& Probabilistic TA & \cellNo{} \\
\hline
\end{tabular}

}
\end{table}

Using timed automata as a target formalism for attack trees is quite recent; in contrast, the theoretical works on opacity surveyed in \cref{section:non-interference} started as early as~2002.
So far, we identified two branches (attack-fault trees and attack-defense trees) whose expressive power differs mainly in terms of the type of the supported logical gates.
Concerning attack-fault trees, \cite{KS17} models stochastic behaviors but does not perform parameter synthesis, while~\cite{ALRS21} performs parameter synthesis, but does not support probabilistic automata.
On the topic of attack-defense trees, \cite{HKKS16} models stochastic behaviors in the more expressive attack-defense diagrams.
On the other hand, \cite{ABPPSS20}~performs parameter synthesis and models several agents.
It is worth mentioning that attack-defense trees were also translated to other (non-TA) timed formalisms, notably to continuous time Markov chains (\eg{} \cite{JLM16,LO20}).

Contrarily to works related to non-interference properties (\cref{section:non-interference}), the works surveyed in this section on attack trees are clearly well-supported by existing model checkers.
With one exception, all of the works in \cref{table:summary-attack-trees} use a target model checker, with some reported experiments.
Several of these works also feature an automated translation from the input attack tree into the target formalism (\ie{} TAs or extensions of~TAs).

As a summary, this direction of research seems to be very dynamic, and a clear future work will be to unify the various approaches (featuring faults, defense, agents, parameters, probabilities…)\ into a unique and fully automated translation.

\section{Applications in the context of cyber-physical systems}
\label{section:applications}

In this section, we review works where timed automata or their extensions are used to solve concrete problems related to security, and that were not covered by the previous two sections of this survey.

\paragraph{Outline of the section}
We first review works aiming at formalizing timed extensions of RBAC models (\cref{ss:RBAC});
we then review works related to the verification of controllers in a security context (\cref{ss:controllers});
we then review a large line of quite different works related to security and communication protocols (\cref{ss:protocols});
we finally review some remaining works (\cref{ss:others}),
and we propose some discussion on these works (\cref{ss:applications:discussion}).

\subsection{Analyses of RBAC models}\label{ss:RBAC}

Several works performed analyses of role-based access control (RBAC) models using timed automata.

In~\cite{MS08}, Mondal and Sural analyze a TRBAC model~\cite{BBF01}, the timed version of a Role Based Access Control model~\cite{SCFY96}.
An RBAC system is composed of three components (role, permission and user).
A formal translation from TRBAC to TAs is provided, where each role is modeled by an automaton, along with a controller automaton that manages the activation/deactivation of these roles.
In order to reduce the size of the state space, the model features only one clock.
This leads to a fast verification time, but the resulting model is a high level abstraction of a real TRBAC.
Security properties are checked using CTL queries with \uppaal{}, ensuring the correct behavior of the system.

In~\cite{MSA11}, Mondal \etal{} extend~\cite{MS08} to GTRBAC~\cite{JBLG05}, an extension of TRBAC allowing to differentiate between enabled role (ready for user assignment) and active role (assumed by at least one user).
Also, a given user can take on multiple roles.
They provide a framework such that a given GTRBAC system is automatically translated into a timed automata network, and CTL queries are generated according to a set of desired properties.
The resulting network of timed automata is proved to possess all the properties of the GTRBAC model that the authors are interested in.
To face the state space explosion issue, the authors provide some abstractions that may be useful to reduce the size of the model, depending on the property to verify.
Experiments with \uppaal{} are used to illustrate the impact of those abstractions on the size of the state space.

In~\cite{GBO12}, Geepalla \etal{} use model checking to analyze spatio-temporal RBAC~\cite{RT08} policies.
Unlike GTRBAC models, those allow to model both temporal and \emph{spatial} aspects of the system.
As a proof of concept, the authors model a banking application with timed automata. %
A property of separation of duty between role (\ie{} that a given user cannot be assigned two exclusive roles at the same time) is verified, and a trace of execution that violates such a property is given.
The system is quite small and scalability is not discussed.
The translation is performed manually---automating the process being left as future work. %

In~\cite{VNN17}, Vasilikos \etal{} propose a method for checking time dependent information flow on distributed systems, typically for RBAC models.
The authors describe a branching-time logic extending CTL, called BTCTL, allowing comparisons between sets of variables, as well as checking if a configuration of variables holds before or after a given event.
Given a distributed system modeled with TAs and queries expressed using BTCTL, an algorithm is provided that syntactically modifies the model and queries so that they can be expressed in the subset of CTL supported by \uppaal{}.
Execution traces are proved to be equivalent.
A translator from BTCTL to the \uppaal{} CTL fragment has been implemented by the authors, but no experiments are reported.

\subsection{Verification of controllers}\label{ss:controllers}

In~\cite{Acampora10}, Acampora introduces TA-based Fuzzy Controllers (TAFC), a kind of fuzzy controllers~\cite{Sugeno85} modeled by timed automata extended with fuzzy variables.
The controller allows for the detection of network intrusions, and it is built based on data collected from routers from the network.
The process is experimented on a DoS (denial of service) attack, and the generated controller is then compared to a classic, untimed fuzzy controller.
The results show that the TAFC makes better prediction than its untimed counterpart.

In~\cite{WSC17}, Wang \etal{} study a networked water level control system with both numerical and physical aspects.
The different parts of the systems are modeled with TAs, along with the controller of the system, and an attacker, each in the form of an automaton.
Some security properties are expressed using LTL and checked with the PAT model checker~\cite{LSD11}.
The results highlight that the system is secure as long as the frequency of the monitoring of the system is greater than or equal to the frequency of the attacks.
The method is very \emph{ad hoc}, and the paper is therefore rather a case study paper.

In~\cite{AHM19}, Alshalalfah \etal{} study the resilience of controllers of artificial pancreas.
Those devices allow for an automated regulation of glucose level.
The system features wireless communication between an insulin pump and a controller that makes decisions about insulin injection.
The components of the system are modeled with priced timed automata, including an attacker monitoring the actions of the controller.
The attacker can perform a replay attack by intercepting messages from the controller and re-sending them later, disturbing the actions of the controller.
Two controllers from the literature are modeled, and stochastic properties are verified with \uppaal{}-SMC.
The first studied controller shows that when attacked, it is still able to maintain safe glucose level in most scenarios, which is not the case for the second one.

\subsection{Verification of security protocols}\label{ss:protocols}

Checking formally the correctness of security protocols is an important yet difficult matter (see \eg{} \cite{KKS16}).
In addition, verification can be even more challenging in the presence of \emph{timing} information.
We review here works related to the verification of security protocols using (extensions of) timed automata.
\label{newtext:SP}

\paragraph{\cite{CEHM04} and extensions}

A method for modeling security protocols via a network of timed automata is proposed for the first time by Corin \etal{} in~\cite{CEHM04,CEHM07}.
The resulting model considers the timed aspects of such protocols, allowing to model not only timeouts, but also the possible actions following a timeout, such as the retransmission of a message.
The behavior of each participant, including the intruder, is modeled by an automaton.
The formal verification of security properties is done using \uppaal{}.
An example is given with a simplified version of the Needham-Schroeder protocol~\cite{NS78,L95} where a Dolev-Yao intruder~\cite{DY83} successfully intercepts the message if the responder timeout is long enough.
The method is then applied to prove that retransmissions are secure on the Yahalom protocol~\cite{BAN90}.
Only protocols requiring timeouts are considered.

In~\cite{JPS05}, Jakubowska \etal{} propose a method similar to~\cite{CEHM04} for modeling security protocols.
Timed automata (extended with global variables) are used to model security protocol such that each participant is modeled by an automaton.
Unlike~\cite{CEHM04}, this work gives a more formal way of specifying security protocols.
While this method does not consider timeout, it models timestamps~\cite{NS93} on the other hand.
Authentication and integrity properties can be checked either using partition refinement with Kronos or SAT bounded model checking with \Verics{}, which may return traces of execution leading to states violating such properties.
The method is then applied to the Kerberos protocol~\cite{BP98} and to the Wide Mouthed Frog protocol (WMF)~\cite{BAN90}, and identifies a violation of authentication in both cases.
Model checking is used to prove that such an attack is not possible anymore when testing freshness of the session keys.
In~\cite{JP07}, Jakubowska and Penczek extend~\cite{JPS05} in various ways.
Timeouts are now considered and timed aspects of timing attacks are studied on the Andrew Secure RPC protocol~\cite{S89}, the Telecommunications Management Network protocol~\cite{SOB88} and the Needham-Schroeder protocol~\cite{NS78}.
In particular, its is described how well-chosen timeout values can prevent an attack on the Needham-Schroeder Public-Key (NSPK) that was described in~\cite{L95}.

In~\cite{BCP09}, Benerecetti \etal{} propose a framework, named TPMC, for checking security properties of timed security protocols.
TMPC extends the specification language HLPSL~\cite{ABB05}, an established language in the security protocol community.
A timed version of HLPSL (namely ``THLPSL'') is used to model protocols, and an automatic translation to \uppaal{} TAs is provided.
The method is exemplified on the Wide Mouthed Frog protocol.
The way systems are modeled and verified is quite similar to the work by Corin \etal{}~\cite{CEHM07}, with the advantage of proposing a formalism that might be easier to use for the security community.\label{oldtext::HLPSL}

In~\cite{KP09}, Kurkowski and Penczek use bounded model checking to verify security protocols modeled using TAs.
(Bounded model checking for (extensions of) timed automata was considered in, \eg{} \cite{KP12BMC,KJN12,WZZ17}.)
This work is relatively close to~\cite{CEHM04}, and extends to \emph{timed} automata a previous work by the same authors on finite-state automata~\cite{KP07}.
In~\cite{KP09}, a formal specification language for timed security protocols is thoroughly described.
Knowledge of participants is part of the model, and attacks correspond to states of the system where an intruder possesses certain knowledge.
A translation to a network of TAs is then proposed.
Checking is done using SAT-based bounded-model checking with \Verics{}.
The method is experimented on several types of security protocols, including WMF, Kerberos and NSPK and is able to capture known attacks on authentication and secrecy.
The computation times are compared to the TMPC framework and show better overall results.

\paragraph{SAT- and SMT-based verification}\label{paragraph:Szymoniak}
We review here a line of works by Szymoniak, Kurkowski and co-authors.
First, in~\cite{SSK16}, the objective is to analyze execution times, and to exhibit a time (``timeout'') during which a protocol can be executed.
An implementation has been carried out, allowing to ``synthesize'' a secure execution time for a protocol.
Experiments were performed on a modification of the Needham Schroeder authentication protocol~\cite{NS78}.
This work is not TA-based; but it shares similarities with the TA-based work by André \etal{}~\cite{ALMS22}, %
	where the authors also synthesize timing parameter valuations (such as timeout) to provide secure execution times for a system.
\label{oldtext:SSK16}

In~\cite{SSK18}, Szymoniak \etal{} apply their method from~\cite{SSK16} to the WooLamPi protocol~\cite{WL94}.
Models of timed automata are generated using a compiler from their previous work and then translated to a Boolean formula to be verified using the SAT-solving tool MiniSAT~\cite{SE05}.
This allows to find a dependency between latency values that forbid man-in-the-middle attack.

Then, in~\cite{ZSK19}, Zbrzezny \etal{} propose a bounded model checking approach based on \emph{Satisfiability Modulo Theory} (SMT).
The underlying model is networks of communicating timed automata, and the modeling is inspired by~\cite{KP12book,SKP15}.
The experiments in~\cite{ZSK19} reuse the properties detailed in~\cite{SSK16}.
Experimental results on several common protocols (NSPK~\cite{NS78}, a timed version of Needham-Schroeder Symmetric Key (NSSK)~\cite[Section~2]{ZSK19}, WooLamPi (WLP)~\cite{WL94} and WMF~\cite{BAN90})\ show that the SMT approach is faster than the previous SAT approach.
In particular, their implementation on the SMT-solver \Yices{}~\cite{Dutertre14} is at least ten times faster than its SAT counterpart (implemented on the SAT-solver MiniSAT).
An artifact for reproducibility is given.

This former work~\cite{ZSK19} is the partial basis of VerSecTis~\cite{ZZSSK20}, a tool dedicated to the verification of timed security protocols with SMT-based techniques.
Here, protocols are modeled as \emph{Timed Interpreted Systems} (TIS)~\cite{WZ16,ZZ17}, instead of network of timed automata.
Again, instructions to reproduce experiments are available online.

In~\cite{SSZZK21}, Szymoniak \etal{} build on previous works~\cite{JP07,KP12book}, and perform a formal modeling and timed analyses of the WooLamPi protocol and of the Sensor Network Encryption Protocol~\cite{PSTWC02}.
Reachability using SMT is used to identify the time frames during which a man-in-the middle behavior (\ie{} a passive interception of the data) is possible.

This line of work is continued into~\cite{Szymoniak21}, where simulations are used to test the influence of time parameters on the correctness of security protocols.
Timed automata are however not used as the underlying model.

\paragraph{Other works}

In~\cite{KKO10}, Koltuksuz \etal{} describe how to model and verify security protocols using TAs.
The method shares similarities with~\cite{CEHM07}, each participant being modeled as an automaton.
An example is given using the Neuman-Stubblebine repeated authentication protocol~\cite{NS93} with a Dolev-Yao intruder; an extensive verification is performed using \uppaal{}.

In~\cite{SKP15}, Szymoniak \etal{} propose a method for modeling security protocols with a particular focus on the \emph{latency} in the network.
Beyond the specificity of the networks delays, the approach is quite classical, describing honest users and intruders behaving on a shared network, modeled with~TAs.
There are no experimental results, and the latency of the communications is modeled by adding some fixed value to the timed constraints of automata.

In~\cite{PG15,PG16}, Piech and Grodzki address the problem of dynamic security estimation, and aim at providing algorithms that detect threats and modify attributes of a communication protocol in real time to adapt to the risks.
To do so, they model a given communication protocol using a probabilistic timed automaton, such that each location in the automaton corresponds to a security state (\ie{} a value modeling the security attributes of the system).
However, this TA is only used as a means of simulating the execution of a given communication protocol, in order to show the results of the provided algorithm.
In~\cite{PG16}, Petri nets are used as an additional formalisms; experiments are performed, but it remains unclear using which framework.

In~\cite{Siedlecka20}, Siedlecka models security protocols with probabilistic timed automata.
The paper extends~\cite{SKP16}, a previous work where security protocols were modeled with (untimed) probabilistic automata.
This work considers both the timed aspects of the protocols (such as the timeframe of tasks or the communication delays), and the probabilistic aspects inherent to the success of the attacks.
The framework allows to know the probability of success of the attacker at a given step of the protocol.
A tool has been implemented and is experimented on the security part of the MobInfoSec protocol~\cite{HPEMCS14} as well as on the NSPKL protocol~\cite{Lowe96}.
Finally, in~\cite{LSLSD18}, Li \etal{} extend the formalism of applied $\pi$-calculus~\cite{ABF18} with parametric timed constraints.
Using this formalism, the authors model security protocols along with an adversary, and they synthesize parameter valuations for which the system is secure.
As the expressiveness of such modeling seems close to that of parametric timed automata, it suffers from the same drawbacks: their algorithms offer no guarantee of termination.
Furthermore, timestamps can be expressed in this formalism, but not timeouts.
The results have been implemented in a tool named Security Protocol Analyzer, which performs an on-the-fly verification of the state space, stopping if no parameter valuations satisfy the property, or whenever the entire state space has been explored.
The tool is run on several security protocols, which finds a previously undiscovered time related attack on Kerberos V (\ie{} the latest version of this protocol).
In~\cite{LSD16}, the authors use the same method to consider clock drifts (\ie{} inaccuracies in clock rates).
They show that the TESLA protocol~\cite{PCST01} is safe when assuming precise timed constraints, but might not be once clock drifts occur.
These two works~\cite{LSD16,LSLSD18} do not consider timed automata, but the methods used share similarities with parameter synthesis for parametric timed automata.
\label{newtext:LSLSD18-LSD16}

\paragraph{Case studies}

In~\cite{GG06}, Godskesen and Gryn briefly report on the formal verification of the authenticated routing for the \emph{ad hoc} networks (ARAN) protocol~\cite{SLDLSB05}.
A TA model of the network, including an attacker, has been implemented and verified with \uppaal{}, allowing to confirm an already known attack along with the (also known) correction needed to prevent it.
No model, queries nor results are shown.

In~\cite{TCCP08}, Tobarra \etal{} describe an analysis of routing protocols in wireless sensor networks~\cite{ASSC02}.
In this type of infrastructure, an intruder can intercept or even modify messages between nodes of the network.
The attacker approach is used on the $\mu$-Tesla authentication protocol~\cite{PSTWC02}, which is modeled with timed automata.
However, the intruder is not modeled with an automaton but with variables denoting its knowledge, along with functions allowing it to increase its own knowledge on regard of the information it already have access to.
Checking is claimed to have been done with \uppaal{}, but no queries nor results are provided.

In~\cite{ADMM14}, Andrychowicz \etal{} model and verify security of Bitcoin contracts.
Two contract protocols from~\cite{ADMM16} are verified: the timed commitment scheme and the simultaneous commitment scheme.
A model is given in timed automata for the blockchain algorithm and two users.
Both users are in turn considered as dishonest.
The resulting network is checked with \uppaal{} and conditions for validation of security properties are given (typically the respect of some maximum time limit before timeout).

In~\cite{SR16}, Saffarian and~Rafe apply model checking using \uppaal{} and timed automata on Mobile Internet Protocol version 6 (MIPv6)~\cite{PJA11}.
Through some level of abstraction, the protocol is modeled, including its real time aspects.
Aside from participants, the network and a Dolev-Yao intruder are modeled by one automaton figuring the unprotected communication media.
Formal verification cover the security of MIPv6 route optimization, and identifies already known attacks scenarios, along with a new one that was not studied previously.

In~\cite{LS18}, Lu and~Sun study the 4-way handshake protocol of the IEEE~802.11i standard.
This kind of protocol is used for authentication between two parties.
Each participant, including an attacker trying to perturb communications with a denial of service attack, is modeled with a~TA.
Using \uppaal{}, it is formally proved that the key generated during the authentication will be received as planned.
However, the specification of the system implies a high level of abstraction which may impact its realism.
\paragraph{Discussion}\label{paragraph:discussion_protocols}
In general, security protocols are modeled with networks of timed automata such that each participant or component is modeled by one automaton.
Among the surveyed frameworks, \cite{Siedlecka20} is the only framework coping with both timed and probabilistic aspects.
Probabilities set aside, \cite{ZSK19} is among the most expressive as it considers timeouts and timestamps, and models the knowledge of the agents. %
Still, it is worth mentioning the framework TPMC from~\cite{BCP09}, as it features a specification language that was designed to be easy to use for the security community.

Finally, none of the works consider a \emph{parameterized} number of participants.
In other words, it may happen that some protocols were proved correct for a given number of participants (two, typically), but some larger number of participants might introduce security breaches.

\subsection{Other works}\label{ss:others}

We finally review some miscellaneous works that did not fit in the structure of the survey.

In~\cite{BMC12}, Burmester \etal{} study security in cyber-physical systems, with multiple attack vectors, \ie{} both on the discrete (``cyber'') and on the continuous (``physical'') aspect of the system.
The system is modeled with hybrid automata~\cite{Henzinger96}, a formalism which is commonly used when it comes to modeling physical aspects of a system; TAs can be seen as a subclass of hybrid automata.
In this work, the intruder is based on the Byzantine faults model~\cite{Dolev82}, where the adversary controls the communication channels of the system.
The considered case study is the Russia-Ukraine natural gas grid, a communication network managing the flow of gas in a grid linking Russia to Europe.
A flow verification protocol is given that allows Ukraine to verify the correctness of flows while guaranteeing privacy of Russia's information.
The security properties are formally proved using existing cryptographic methodologies.

In~\cite{SWMC15}, Sun \etal{} apply model checking techniques in the context of the Internet of Things, specifically Electronic Product Code (EPC) networks~\cite{SBE01}.
They model a supply chain management system using a timed automata network.
Various potential attacks are modeled, mainly related to Radio Frequency IDentification (RFID) tags.
The verification is performed with \uppaal{} on a concrete system instance, and establishes a safe refresh frequency for the detection of tags.

In~\cite{DDHLLOP15}, David \etal{} study socio-technical attacks using formal verification.
In particular, a case study of the home payment service of an internet TV service is presented.
This service allows the user's payment card to be read by a card reader built into the TV remote.
Typically, an attacker can trick the user to install malicious software.
The authors use TAs to model the physical actions possible for the attacker (such as scamming or physically intruding) and for the user (such as enabling a localization or losing the credit card).
\uppaal{} is used to perform formal verification, and several attacks scenarios are captured that way.
\subsection{Discussion}\label{ss:applications:discussion}

The works surveyed in this section cover multiple application domains.
Many of the aforementioned verification techniques reduce to reachability checking, and therefore \uppaal{} is often the model checker of choice.
In addition, in the last three decades, numerous works addressed the \emph{safety} of protocols, by modeling these protocols using timed automata, and carrying automated verification, often with \uppaal{}; it is therefore not surprising that the \emph{security} of protocols is an application domain of predilection for timed automata.

In terms of concrete application domains, many works addressed application domains related to networks and communications.
This is unsurprising considering that security issues are mainly considered in communicating contexts, when intruders can observe or interfere with communications between legitimate stakeholders.
More surprising is the very small number of works (mainly~\cite{AHM19}) targeting health applications.
Digital devices used in a health environment are typical examples of cyber-physical systems; in addition, their functional correctness and security should be of utmost importance considering their safety-critical nature.
One may wonder why so few works were performed in this area: perhaps these communities are not so interested in formal verification (which would be worrying); perhaps the systems are too expressive for being modeled by timed automata (\eg{} continuous variables more expressive than simple TA clocks might be necessary); perhaps those systems are too large to be verified by model checking and their designers failed in designing well-balanced abstractions.

\section{Conclusions}\label{section:conclusions}

In this manuscript, we surveyed works using timed automata as a formalism for expressing, studying or formally verifying various categories of security properties.
We notably collected a number of works focusing on theoretical aspects of non-interference and opacity properties, on attack trees and their extensions, on RBAC models, and on various security protocols.
In the following, we draw some high-level conclusions and discuss perspectives.

\subsection{Theoretical issues}
The discussed theoretical aspects mostly concern opacity and non-interference.
They naturally collide with the known (un)decidability results for TAs and their subclasses such as event-recording automata (ERAs).
It seems that the undecidability of the timed language inclusion in timed automata is a major point in the study of opacity in timed automata.
While timed opacity is undecidable for the language for the full class of TAs (and even for ERAs)~\cite{Cassez09}, decidability can be envisioned by weakening the attacker model (that could only measure the full execution time~\cite{ALMS22}), by restraining the formalism~\cite{WZA18}, or by restraining the problem by adding a time bound~\cite{AEYM21}---to go back to the decidability of time-bounded language inclusion checking for TAs.

In addition, extending these problems to \emph{parametric timed} extensions of TAs, such as the upper-bound automata (U-PTAs) or lower-bound automata (L-PTAs)~\cite{BlT09}, known for their partially decidable properties (\eg{} \cite{BlT09,ALR18FORMATS}), seems an interesting future work.

\subsection{Links with applications}
\paragraph{Modeling with TAs}
Among the theoretical works on security properties for TAs (mostly surveyed in \cref{section:non-interference}), a blind spot seems to be the relationship with concrete applications.
Not only few of the surveyed works are equipped with implementations, but the relationship with more concrete or higher-level formalisms is virtually not discussed.
That is, while timed automata represent an interesting formalism, and while studying security properties on this formalism is of interest by itself, nevertheless relating these models and properties with actual systems (programming languages, protocols…)\ was not discussed at all in the works surveyed in \cref{section:non-interference}.
This is even more frustrating as
\cref{section:attacktrees,section:applications} convince us that TAs are a suitable formalism to verify security properties in actual systems.

Translations from applications (such as code) to timed automata would be highly interesting, so as to use model checking techniques for TAs as a way to prove opacity of actual programs---or other areas of applications.

\paragraph{Successful research areas}
Perhaps surprisingly, two research directions were thoroughly studied (and are still ongoing), namely the translation of attack trees and their extensions, as well as of RBAC models, to (extensions of) timed automata.
These works are often complete, in the sense that the input formalism is usually completely supported, and most works come with an automated translation, and a target model checker supporting (extensions of) TAs.

On the opposite, health applications are surprisingly virtually absent from the surveyed works (see discussion in \cref{ss:applications:discussion}).

\paragraph{Extensions}
We mostly limited ourselves to ``vanilla'' timed automata, and considered few extensions (mostly probabilistic, and parametric), mainly because they appeared reasonably frequently in the surveyed works.
Other formalisms of interest, potentially used in the context of formally assessing security properties, include hybrid automata~\cite{Henzinger96}, or continuous time Markov chains~\cite{LO20}.

\paragraph{Tool support}
\uppaal{}, which is one of the most popular tools for the verification of TAs, is the most widely used tool among the applications surveyed in \cref{section:applications}, which can be seen as unsurprising.
More surprising, \uppaal{} is not much used in the context of attack trees, perhaps due to the need of probabilistic and parametric extensions needed by the surveyed attack tree frameworks.
Also surprising at first, \uppaal{} is not much used in the context of non-interference and opacity, mainly because these notions are close to \emph{language inclusion}, a problem undecidable for TAs, while \uppaal{} mainly focuses on reachability.

We believe the community should develop new tools efficient at language inclusion for either decidable subclasses (such as event-recording automata, or one-clock automata), or for the whole class of TAs, with the risk that the algorithms do potentially not terminate (this is notably the paradigm of the algorithm \arXivVersion{and the tool }proposed in~\cite{WSLWL14}, with a very frequent termination in practice).

\paragraph{Tool reliability}\label{paragraph:reliability}
A discussion quite absent from the surveyed works is that of the tool \emph{reliability}, either because of tool failures (due to bugs), or due to modeling issues.
For example, in~\cite{KKS16}, Kurkowski et al.\ showed that some incorrect protocols can be verified positively by some tools (\eg{} TA4SP).
Modeling issues can involve Zeno behaviors, \ie{} an infinite number of discrete actions in a finite time, or deadlocks or timelocks.
While techniques exist to avoid these problems in timed automata (\eg{} \cite{GB07,WSWLSDYL15}), maybe not all surveyed works considered these issues.
In addition, the guaranteed absence of bugs in timed automata model checkers remains a real challenge.
Only few recent works attempted at a certified verification, notably by Wimmer \etal{} and the Munta verified model checker for timed automata~\cite{WL18,Wimmer19,WM20,WHP20}.
These works do not seem to have been used for proving security protocols.

\paragraph{Reproducibility}\label{paragraph:reproducibility}
The surveyed works describing implementations and experiments do generally not provide their code, models and experiment results on long-term archiving venues.
This may result in loss of data in a couple of years.
This is not specific to this area of research, but is even more unfortunate, as most data could be archived very easily (usually no sensitive data, reasonably-sized models and results, etc.).
Fortunately, some recent exceptions \emph{do} provide online benchmarks with all necessary information for reproducibility (\eg{} \cite{ZSK19,ZZSSK20,ALMS22}).

\subsection{Towards parameterized verification using timed automata?}\label{ss:parameterized}
A research domain notably absent of our survey on timed automata is that of \emph{parameterized} models.
Whereas some works considered \emph{timing parameters} (\eg{} \cite{LSLSD18,AK20,ABPPSS20,ALMS22}), \emph{parameterized} models in terms of number of participants, are absent from the surveyed works.
In other words, it may happen that some protocols surveyed in this manuscript were proved correct for a fixed number of participants, but could turn insecure for larger number of participants.

This problem was tackled in various areas of security---but, to the best of our knowledge, not using timed automata.
This problem was for example tackled in~\cite{CC04} for a class of (untimed) protocols, by projecting all honest users onto one agent and all attackers onto another one---it is therefore sufficient to consider only two~agents.
Achieving something similar in timed automata is far from trivial as, depending on the modeling choices, we might need to equip each agent with its own clock, which creates an unbounded number of clocks.

Nevertheless, there is a relatively long line of works studying the decidability of ``timed network protocols'', \ie{} networks made of an arbitrary number of timed automata, communicating with various paradigms and restricted by various constraints, such as the number of clocks per agent (\eg{} \cite{AJ03,BF13,ADRST16}).
An explanation why these results have not been used in the context of security may come from the fact that virtually none of these works led to an implementation.

\ifdefined\VersionAuthorforArXiV
\else
	\begin{acks}
		\ouracks{}
	\end{acks}
\fi
\ifdefined\VersionAuthorforArXiV
	\newcommand{\AISC}{Advances in Intelligent Systems and Computing}
	\newcommand{\LNCS}{Lecture Notes in Computer Science}
\else
	\newcommand{\AISC}{AISC}
	\newcommand{\LNCS}{LNCS}
\fi

\newcommand{\CCIS}{Communications in Computer and Information Science}
\newcommand{\ENTCS}{Electronic Notes in Theoretical Computer Science}
\newcommand{\FAC}{Formal Aspects of Computing}
\newcommand{\FundInf}{Fundamenta Informaticae}
\newcommand{\FMSD}{Formal Methods in System Design}
\newcommand{\IJFCS}{International Journal of Foundations of Computer Science}
\newcommand{\IJSSE}{International Journal of Secure Software Engineering}
\newcommand{\IPL}{Information Processing Letters}
\newcommand{\JAIR}{Journal of Artificial Intelligence Research}
\newcommand{\JLAP}{Journal of Logic and Algebraic Programming}
\newcommand{\JLAMP}{Journal of Logical and Algebraic Methods in Programming} %
\newcommand{\JLC}{Journal of Logic and Computation}
\newcommand{\LMCS}{Logical Methods in Computer Science}
\newcommand{\RESS}{Reliability Engineering \& System Safety}
\newcommand{\STTT}{International Journal on Software Tools for Technology Transfer}
\newcommand{\TCS}{Theoretical Computer Science}
\newcommand{\ToPNoC}{Transactions on Petri Nets and Other Models of Concurrency}
\newcommand{\TSE}{{IEEE} Transactions on Software Engineering}

\ifdefined\VersionAuthorforArXiV
	\renewcommand*{\bibfont}{\footnotesize}
	\printbibliography[title={References}]
\else
	\bibliographystyle{ACM-Reference-Format} %
	\bibliography{survey}
\fi

\end{document}